\documentclass[prl,a4paper,twocolumn,showpacs,floatfix,superscriptaddress,amsmath,amsfonts,amssymb,preprintnumbers,citeautoscript]{revtex4-1}
\pdfoutput=1
\usepackage[T1]{fontenc}\usepackage[latin1]{inputenc}
\usepackage{dcolumn,graphicx,color,booktabs,microtype,afterpage} \graphicspath{{Figures/}}
\usepackage[autoplay]{animate}
\usepackage[charter,greekuppercase=italicized]{mathdesign}\usepackage{sidecap}
\renewcommand{\figurename}{Fig.}
\renewcommand{\tablename}{Table}
\makeatletter\renewcommand{\fnum@figure}[1]{\figurename~\thefigure~(color online). }\makeatother
\makeatletter\renewcommand{\fnum@table}[1]{\tablename~\thetable.}\makeatother

\newcount\hh \newcount\mm
\hh=\time \divide\hh by 60
\mm=\hh \multiply\mm by 60 \mm=-\mm
\advance\mm by \time
\def\now{\number\hh:\ifnum\mm<10{}0\fi\number\mm}

\usepackage[colorlinks,plainpages=false,linkcolor=blue,urlcolor=blue,citecolor=blue,pdfpagemode=UseNone,pdfstartview=FitBH]{hyperref}

\begin{document}

\makeatletter\renewcommand{\ps@plain}{%
\def\@evenhead{\hfill\itshape\rightmark}%
\def\@oddhead{\itshape\leftmark\hfill}%
\renewcommand{\@evenfoot}{\hfill\small{--~\thepage~--}\hfill}%
\renewcommand{\@oddfoot}{\hfill\small{--~\thepage~--}\hfill}%
}\makeatother\pagestyle{plain}


\title{Non-collinear antiferromagnetism of coupled spins\\and pseudospins in the double perovskite La$_\text{2}$CuIrO$_\text{6}$}

\author{Kaustuv\,Manna}\email[Corresponding author: \vspace{4pt}]{kaustuvmanna@gmail.com}

\altaffiliation{\\Presently at the Max Planck Institute for Chemical Physics of Solids, N\"othnitzer Stra\ss e-40, 01187 Dresden, Germany \vspace{8pt}}

\affiliation{Leibniz-Institute for Solid State and Materials Research, Helmholtzstra\ss e 20, 01069 Dresden, Germany}

\author{R.\,\,Sarkar}
\affiliation{Institut f{\"u}r Festk{\"o}rperphysik, TU Dresden, D-01069 Dresden, Germany}

\author{S.\,Fuchs}
\affiliation{Leibniz-Institute for Solid State and Materials Research, Helmholtzstra\ss e 20, 01069 Dresden, Germany}

\author{Y.\,\,A.\,\,Onykiienko}
\affiliation{Institut f{\"u}r Festk{\"o}rperphysik, TU Dresden, D-01069 Dresden, Germany}

\author{A.\,K.\,\,Bera}
\affiliation{Solid State Physics Division, Bhabha Atomic Research Centre, Mumbai 400 085, India}

\author{G.\,Aslan\,\,Cansever}
\affiliation{Leibniz-Institute for Solid State and Materials Research, Helmholtzstra\ss e 20, 01069 Dresden, Germany}

\author{S.\,\,Kamusella}
\affiliation{Institut f{\"u}r Festk{\"o}rperphysik, TU Dresden, D-01069 Dresden, Germany}

\author{A.\,\,Maljuk}
\affiliation{Leibniz-Institute for Solid State and Materials Research, Helmholtzstra\ss e 20, 01069 Dresden, Germany}

\author{C.\,G.\,\,F.\,Blum}
\affiliation{Leibniz-Institute for Solid State and Materials Research, Helmholtzstra\ss e 20, 01069 Dresden, Germany}

\author{L.\,T.\,\,Corredor}
\affiliation{Leibniz-Institute for Solid State and Materials Research, Helmholtzstra\ss e 20, 01069 Dresden, Germany}

\author{A.\,U.\,B.\,Wolter}
\affiliation{Leibniz-Institute for Solid State and Materials Research, Helmholtzstra\ss e 20, 01069 Dresden, Germany}

\author{S.\,M.\,\,Yusuf}
\affiliation{Solid State Physics Division, Bhabha Atomic Research Centre, Mumbai 400 085, India}

\author{M.\,Frontzek}
\affiliation{Laboratory for Neutron Scattering and Imaging, Paul Scherrer Institut, CH-5232 Villigen PSI, Switzerland}
\affiliation{Quantum Condensed Matter Division, Oak Ridge National Laboratory (ORNL), Oak Ridge, TN 37831-6475, USA}

\author{L.\,Keller}
\affiliation{Laboratory for Neutron Scattering and Imaging, Paul Scherrer Institut, CH-5232 Villigen PSI, Switzerland}

\author{M.\,Iakovleva}
\affiliation{Leibniz-Institute for Solid State and Materials Research, Helmholtzstra\ss e 20, 01069 Dresden, Germany}
\affiliation{Kazan E. K. Zavoisky Physical Technical Institute of RAS, 420029 Kazan, Russia}

\author{E.\,Vavilova}
\affiliation{Kazan E. K. Zavoisky Physical Technical Institute of RAS, 420029 Kazan, Russia}

\author{H.\,-J.\,Grafe}
\affiliation{Leibniz-Institute for Solid State and Materials Research, Helmholtzstra\ss e 20, 01069 Dresden, Germany}

\author{V.\,Kataev}
\affiliation{Leibniz-Institute for Solid State and Materials Research, Helmholtzstra\ss e 20, 01069 Dresden, Germany}

\author{H.-H.~Klauss}
\affiliation{Institut f{\"u}r Festk{\"o}rperphysik, TU Dresden, D-01069 Dresden, Germany}

\author{D.~S.\,\,Inosov}
\affiliation{Institut f{\"u}r Festk{\"o}rperphysik, TU Dresden, D-01069 Dresden, Germany}

\author{S.\,Wurmehl}
\affiliation{Leibniz-Institute for Solid State and Materials Research, Helmholtzstra\ss e 20, 01069 Dresden, Germany}
\affiliation{Institut f{\"u}r Festk{\"o}rperphysik, TU Dresden, D-01069 Dresden, Germany}

\author{B.\,B{\"u}chner}
\affiliation{Leibniz-Institute for Solid State and Materials Research, Helmholtzstra\ss e 20, 01069 Dresden, Germany}
\affiliation{Institut f{\"u}r Festk{\"o}rperphysik, TU Dresden, D-01069 Dresden, Germany}

\begin{abstract}
\noindent We report the structural, magnetic and thermodynamic properties of the double perovskite compound La$_{2}$CuIrO$_{6}$ from X-ray, neutron diffraction, neutron depolarization, \textit{dc} magnetization, \textit{ac} susceptibility, specific heat, muon-spin-relaxation ($\mu$SR), electron-spin-resonance (ESR) and nuclear magnetic resonance (NMR) measurements. Below $\sim 113$ K, short-range spin-spin correlations occur within the Cu$^{2+}$ sublattice. With decreasing temperature, the Ir$^{4+}$ sublattice progressively involves in the correlation process. Below $T = 74$ K, the magnetic sublattices of Cu (spin $\textit{s}$ = $ \frac{1}{2} $) and Ir (pseudospin \textit{j} = $ \frac{1}{2} $) in La$_{2}$CuIrO$_{6}$ are strongly coupled and exhibit an antiferromagnetic phase transition into a non-collinear magnetic structure accompanied by a small uncompensated transverse moment. A weak anomaly in \textit{ac}-susceptibility as well as in the NMR and $\mu$SR spin lattice relaxation rates at 54 K is interpreted as a cooperative ordering of the transverse moments which is influenced by the strong spin-orbit coupled 5\textit{d} ion Ir$^{4+}$. We argue that the rich magnetic behavior observed in La$_{2}$CuIrO$_{6}$ is related to complex magnetic interactions between the strongly correlated spin-only 3\textit{d} ions with the strongly spin-orbit coupled 5\textit{d} transition ions where a combination of the spin-orbit coupling and the low-symmetry of the crystal lattice plays a special role for the spin structure in the magnetically ordered state.
\end{abstract}

\pacs{75.50.Ee, 78.70.Nx, 76.30.-v, 76.60.-k\vspace{-0.7em}}

\maketitle

\vspace{-5pt}\section{Introduction}\vspace{-5pt}

In recent years, iridates have drawn considerable research interest in the condensed matter physics community due to their comparable energy scales of strong spin-orbit coupling (SOC), Coulomb interaction, crystal field, and exchange interactions. Through the change of local environment and symmetry, the electronic structure as well as the magnetic and transport properties can be tuned. By a small rearrangement of these interaction energies, interesting physical properties like: topologically insulating behavior \cite{Mott,2}, superconductivity \cite{3}, Weyl semimetallic state \cite{4}, quantum spin liquid phases \cite{5,6} etc. are predicted in these materials. An interesting scenario appears for the perovskite iridium oxides where the 5$d$ Ir ions are located within octahedral oxygen cages. For systems with Ir$^{4+}$ ions, the strong SOC couples the spins of 5$d$-electrons residing on the $t_\text{2g}$ orbital to its orbital momentum. The resulting doubly degenerate ground state can be described by an effective total angular momentum (or pseudospin) $j$ = $ \frac{1}{2} $ which can have properties very different from the spin-only $s = \frac{1}{2} $ state such as in the 3$d$ Cu$^{2+}$ ions. In the band picture, an overlap of the $j$ pseudospin states gives rise to a $``J_\text{eff} =  \frac{1}{2} "$ Mott insulator state \cite{7}. In fact, the details of this $J_\text{eff}$ picture, and how the non-cubic crystal field can tune the corresponding band structure are still not clear and are under investigation.

Another level of complexity arises when strongly correlated 3$d$ ions with almost completely quenched orbital moment coexist with strongly spin-orbit coupled 5$d$ ions in the B-site of a double perovskite. An interesting physics emerging from such compounds promises to open up new pathways to develop novel materials with desired magnetic functionalities for advanced technological applications. The ground-state magnetic properties are also found to be versatile in these double perovskites. For example, among the La$_{2}B$IrO$_{6}$ compounds, $B$ = Mn is ferromagnetic (FM) \cite{8}, $B$ = Co, Ni, Fe are reported to possess non-collinear magnetism \cite{9,10}, and the $B$ = Zn compound shows canted antiferromagnetism \cite{11}. An interesting magnetic behavior with FM and AFM features is observed for the $B$ = Cu compound \cite{12,13,14}. Here, the magnetic properties are determined by the strong exchange anisotropy induced by the AFM superexchange interactions between the total angular momentum $J_\text{eff}$ = $ \frac{1}{2} $ of the Ir$^{4+}$ ions and electronic spins $s = \frac{1}{2}$ of Cu$^{2+}$ ions. 

In this paper, we report the results of a systematic investigation of the double perovskite La$_{2}$CuIrO$_{6}$ with state-of-the-art magnetometry, neutron diffraction, neutron depolarization, specific heat, muon-spin-relaxation ($\mu$SR), electron-spin-resonance (ESR) and nuclear magnetic resonance (NMR) to unveil its complex magnetic behavior. The magnetometry measurements show an AFM order below $T_{1}\sim 74$\,K, accompanied by the formation of a weak uncompensated magnetization at $T_{2}\approx 54$\,K. A peak in the low field \textit{ac} susceptibility as well as an anomaly in the NMR and $\mu$SR spin lattice relaxation rates are observed at $T_{2}$ as well. The system possesses an open hysteresis loop below $T_{2}$ and a positive Curie-Weiss temperature.  The neutron diffraction and the $\mu$SR studies reveal a commensurate AFM state below 74\,K with a non-collinear spin structure. A comprehensive analysis of specific heat, ESR and neutron depolarization data reveals that the spins of Cu$^{2+}$ and the pseudospins of Ir$^{4+}$ ions become correlated in a short-range ordered state below 113 K giving rise to small dynamically correlated uncompensated moments. We propose a tentative spin structure in the AFM ordered ground state of La$_{2}$CuIrO$_{6}$ and discuss a multifaceted role of the spin-orbit coupling for the magnetism of this material. Our in-depth investigation illustrates a new pathway to understand the complex physics of various 3$d$-5$d$ mixed valence compounds.

\vspace{-10pt}\section{Experimental Details}\vspace{-5pt}

The polycrystalline sample of La$_{2}$CuIrO$_{6}$ was synthesized using a solid-state synthesis method with high purity La$_{2}$O$_{3}$ (Alfa Aesar, 99.99\%), CuO (Alfa Aesar, 99.99\%) and IrO$_{2}$ (Alpha Aesar, 99.99\%) precursors. The La$_{2}$O$_{3}$  was preheated at 900 $^\circ$C for 12 h in air. Then, all ingredients were mixed in their stoichiometric ratio and thoroughly ground using agate pestle and mortar. The mixture was then calcined in air at 900 $^\circ$C for 24 h followed by a cooling of the furnace at a rate of 150 $^\circ$C/h. The residue was further air annealed consecutively at 1000, 1100 and 1150 $^\circ$C for 60 h each, with several intermediate grindings. The phase formation was monitored at every stage of the annealing process by powder x-ray diffraction (XRD). The XRD patterns were collected in transmission geometry with Mo-K$\alpha$1 radiation using a StoeStadi-Powder diffractometer equipped with a Ge(111) primary monochromator and a DectrisMythen 1\,K detector. Data were fitted by the Rietveld method \cite{15} using FullProf in the WinPlotR program package \cite{16}. 

The compositional and microstructural analyses were performed using scanning electron microscopy (SEM) [Zeiss EVOMA15] along with an electron microprobe analyzer for semi-quantitative elemental analysis in the energy dispersive x-ray (EDX) mode (X-MaxN20 detector from Oxford Instruments with a AZtecEnergy Advanced acquisition and EDX analysis software). The stoichiometry was confirmed close to the starting composition. 
The \textit{dc} magnetization measurements were performed in a Quantum Design SQUID magnetometer from room temperature down to 1.8 K with magnetic fields up to 5 T. The \textit{ac} susceptibility and specific heat measurements were performed in a commercial Quantum Design PPMS in the temperature range $1.8 - 300$\,K with maximum \textit{dc} field up to 9 T. For the susceptibility, an \textit{ac} field of 17 Oe was applied with a frequency up to 1 kHz. 

Neutron diffraction measurements were performed at the DMC cold-neutron powder diffractometer of the Paul Scherrer Institute. The powder sample with a total mass of $ \sim  5$ g was placed into an aluminum container in the form of an annular cylinder to reduce neutron absorption by iridium. The sample was then mounted in a standard ``orange''-type cryostat with the base temperature of 1.6 K. 

The ESR measurements were carried out with a homemade high-field ESR spectrometer \cite{17} upgraded with the PNA-X Network Analyzer from Keysight Technologies. Magnetic fields up to 16 T were obtained with a solenoid-superconducting magnet from Oxford Instruments equipped with a $^{4}$He variable-temperature insert. 

The one--dimensional neutron--depolarization measurements were carried out using the polarized neutron spectrometer (PNS) at the Dhruva reactor ($\lambda = 1.205$ \AA), Trombay, Mumbai, India. Here, the sample is first cooled from room temperature down to 2\,K in the presence of a 10\,Oe field (required to maintain the neutron beam polarization at the sample position), then the transmitted neutron beam polarization was measured as a function of sample temperature in warming cycle under the same field.

In NMR measurements, the $^{139}$La spectra were measured in a magnetic field of 7 T (42.0994 MHz $^{139}$La frequency) with a Tecmag Apollo solid-state spectrometer and a cold bore 16 T field-sweep superconducting magnet from Oxford Instruments. The spectra were collected by point-by-point sweeping of the magnetic field and integration of the Hahn spin echo at each field step. The longitudinal and transversal nuclear spin relaxation rates, $T_{\rm l}^{-1}$ and $T_{\rm t}^{-1}$, were measured with stimulated echo and Hahn spin-echo pulse protocols, respectively.

The zero magnetic field $\mu$SR spectra were recorded in a temperature range $5-130$\,K using the GPS instrument at the PSI Villigen, Switzerland.

\begin{table}
\caption{\label{tab:latticeparameters}
The Rietveld refined crystallographic parameters, such as fractional atomic coordinates, isotropic thermal parameters ($B_{\rm iso}$) obtained from the room temperature XRD. ``Occ.'' corresponds to site occupancies. Space group: $\textit{P}\overline{1}$, $\textit{a} = 5.8000(5)$\,\AA,	$\textit{b} = 7.7279(6)$\,\AA,	$\textit{c} =5.6023(4)$\,\AA,	$\alpha = 89.88(1)^\circ$,	$\beta = 92.59(1)^\circ$, $\gamma = 90.16(1)^\circ$.}
\begin{ruledtabular}
\begin{tabular}{cccccccc}
Atom & $x/a$ & $y/b$ & $z/c$ & $B_{\rm iso}$ & Occ. \\
\hline\\[-1.5mm]
La1  & $0.5556(9)$ & $0.2493(14)$ & $0.4905(10)$ & $0.43(6)$ & $1.0$ \\
La2 & $-0.0519(9)$ & $0.7525(14)$ & $-0.0103(10)$ & $0.43(6)$ & $1.0$ \\
Cu1  & $0.5$ & $0$ & $0$  & $0.29(5)$ & $1.0$ \\
Cu2& $0$ & $0.5$ & $0.5$  & $0.29(5)$ & $1.0$ \\
Ir1 & $0$ & $0$ & $0.5$  & $0.77(8)$ & $1.0$ \\
Ir2 & $0.5$ & $0.5$ & $0$  & $0.77(8)$ & $1.0$ \\
O1  & $0.2842(33)$ & $0.0446(33)  $ & $0.2968(30)$  & $1.38(3)$ & $1.0$ \\
O2  & $0.3139(31)$ & $0.4629(38) $ & $0.3062(31)$ & $1.38(3)$ & $1.0$ \\
O3  & $0.2093(29)$ & $-0.0461(32)$ & $0.7887(35)$ & $1.38(3)$ & $1.0$ \\
O4  & $0.1987(28)$ & $0.5357(36)$ & $0.8123(41)$ & $1.38(3)$ & $1.0$ \\
O5  & $0.4743(15)$ & $0.2443(38)$ & $0.9162(13)$ & $1.38(3)$ & $1.0$ \\	 
O6  & $0.0244(14)$ & $0.7546(37)$ & $0.4213(16)$ & $1.38(3)$ & $1.0$\\ 
\hline\\[-1.5mm]
$R_{\rm p}$  		& $5.14\%$	& $R_{\rm exp}$  	& $4.87\%$ &  \\
$R_{\rm wp}$  	& $6.62\% $	& $R_{\rm Bragg}$ 	& $4.91\% $ & $\chi^2$   	& $1.85 $ 	 \\
\end{tabular}
\end{ruledtabular}
\end{table}

\begin{table*}[t]
\caption{\label{tab:bondlengths}
The bond lengths and bond angles of La$_{2}$CuIrO$_{6}$ at room temperature. }
\begin{ruledtabular}
\begin{tabular}{cccccccc}
 & \multicolumn{6} {c} {Bond lengths (\AA)}  & Octahedral distortion \\\hline
 		& O1					& O2	 & O3	 & O4 	& O5 	& O6	 &\\
Cu1	& 2$\times$2.159(18) &  & 2$\times$2.043(19) &  & 2$\times$1.950(30) &   & 	0.001744 \\
Cu2	&  & 2$\times$2.185(18) & & 2$\times$2.070(20) & & 2$\times$2.020(30)  & 	0.00109 \\
Ir1	&  2$\times$2.075(18)&  &2$\times$2.010(20) & & & 2$\times$1.950(30) & 	0.00109 \\

Ir2	&  & 2$\times$2.088(18) & & 2$\times$2.014(19) & 2$\times$2.040(30) & & 	0.00022\\ \\
\multicolumn{8} {c} {$\langle$ La1--O $\rangle$ = 2.560(18) \AA; $\langle$La2--O$\rangle$ = 2.580(20) \AA} \\ \\
\hline
\multicolumn{8} {c} {Bond angles (deg)}  \\
\hline
\multicolumn{4} {c} {In-plane ($ac$ plane)} & \multicolumn{4} {c} {Out-of-plane ($\textit{b}$ axis)}  \\
\multicolumn{4} {c} {Cu1--O1--Ir1 = 154.3(7)} & \multicolumn{4} {c} {Cu1--O5--Ir2 =151.8(12)}  \\
\multicolumn{4} {c} {Cu1--O3--Ir1 = 152.3(8)} & \multicolumn{4} {c} {Cu2--O6--Ir1 =152.6(12)}  \\
\multicolumn{4} {c} {Cu2--O2--Ir2 = 150.1(7)} &  \multicolumn{4} {c} {}  \\
\multicolumn{4} {c} {Cu2--O4--Ir2 = 149.1(8)} &  \multicolumn{4} {c} {}  \\
\end{tabular}
\end{ruledtabular}
\end{table*}

\begin{figure*}[t]
\begin{center}
\includegraphics[width=0.75\textwidth]{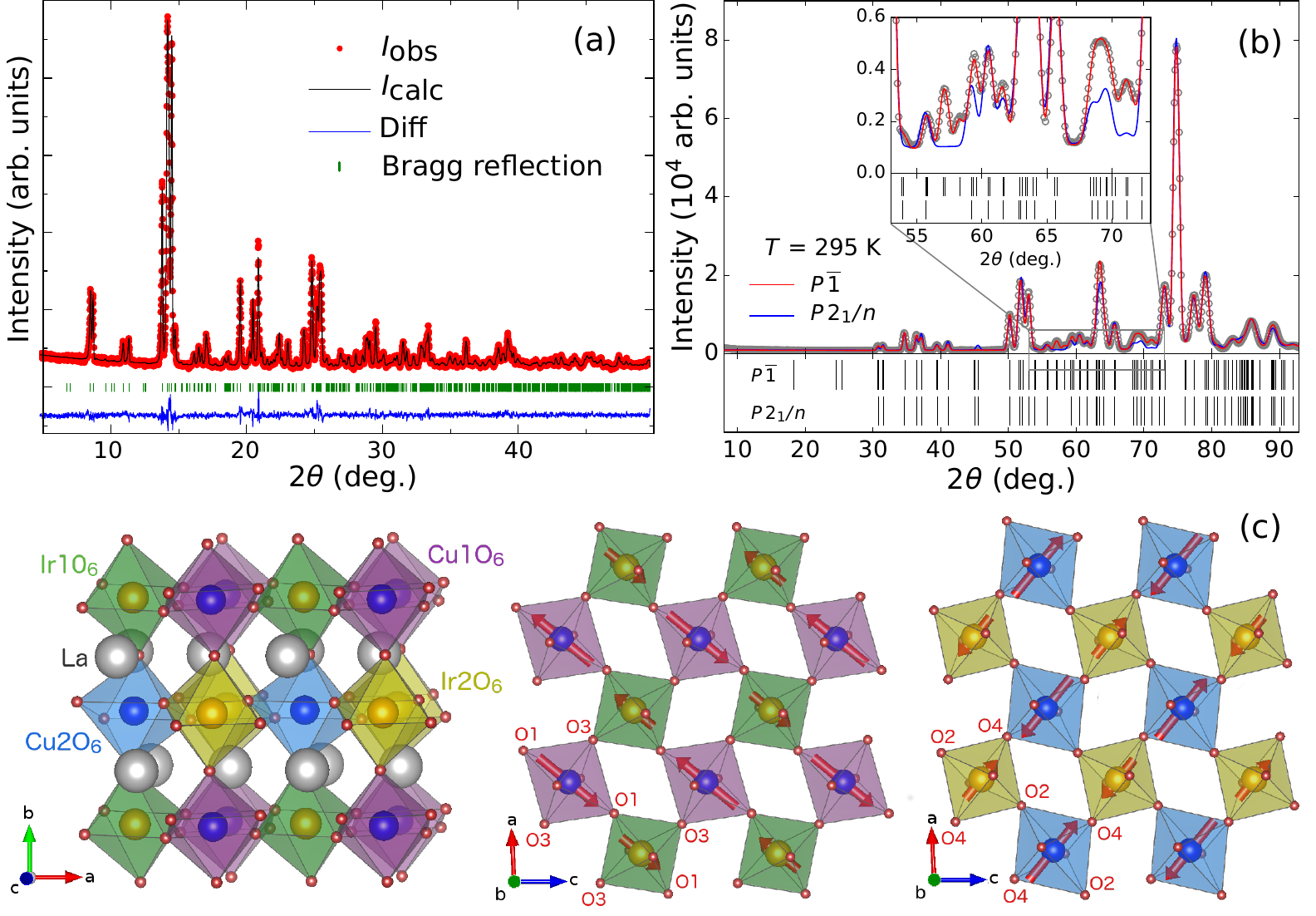}\vspace{-5pt}
\end{center}
\caption{(a) Rietveld refinement fit with triclinic $\textit{P}\overline{1}$ space group of La$_{2}$CuIrO$_{6}$ powder XRD pattern. The red symbol presents the observed intensity, black line is the fitted pattern, blue line is the corresponding difference intensity and the green vertical symbols are the allowed Bragg reflections. (b) Neutron diffraction pattern at 295 K with fitted patterns by both $\textit{P}2_{1}/\textit{n}$ and $\textit{P}\overline{1}$ space groups. The two strongest reflections originate from the aluminum of the sample container (their corresponding marks are not shown). The inset presents a blown-up fragment of the diffraction pattern, where the advantage of the $\textit{P}\overline{1}$ fit over the $\textit{P}2_{1}/\textit{n}$ fit is most clearly seen. (c) Octahedral tilt in three crystallographic directions and the distortion in two types of in-plane octahedral arrangements with [Cu1O$_{6}$ -- Ir1O$_{6}$] and [Cu2O$_{6}$ -- Ir2O$_{6}$] types, as generated from the XRD refinement. The spin configuration is one of the possible patterns compatible with our data obtained from the neutron diffraction refinement.}
\label{fig:XRD}\vspace{-5pt}
\end{figure*}

\vspace{-5pt}\section{Results and Discussion}\vspace{-5pt}

\vspace{-5pt}\subsection{Crystal Structure}\vspace{-5pt}

The room temperature crystal structure of La$_{2}$CuIrO$_{6}$ is determined through an in-depth XRD and neutron diffraction analysis on the polycrystalline sample. The Rietveld refinement reveals that the sample is single phase. The fitted XRD pattern confirms that La$_{2}$CuIrO$_{6}$ crystallizes in the triclinic $\textit{P}\overline{1} $ (No. 2) space group [$\chi ^{2} $= 1.85; $ \textit{R}_{\rm WP} $ = 6.62\%; $ \textit{R}_{\rm P} $ = 5.14\%; $ \textit{R}_{\rm Bragg} $ = 4.91\%.]. An attempt to fit the pattern using the reported space group $\textit{P}2_{1}/\textit{n}$ does not yield a proper fit ($\chi^{2} = 2.12$; $R_\text{WP} = 7.04$\%; $R_{\rm P} = 5.41$\%; $R_{\rm Bragg} = 5.66$\%) \cite{14,18,19}. Moreover, the structural model with the monoclinic $P2_{1}/\textit{n}$ (No. 14) space group could not properly fit the observed neutron diffraction pattern. Specially the Bragg peak at 2$\theta \approx 57.10^{\circ}$ remained unindexed with the $\textit{P}2_{1}/\textit{n}$ (No. 14) space group. The commensurate position of this Bragg peak with respect to the monoclinic structure suggests lowering of the structural symmetry. The observed neutron diffraction pattern could be well described only by the lower-symmetry triclinic $\textit{P}\overline{1} $ (No. 2) space group in consistent with the powder XRD data measured on the same sample. The Rietveld refined fitting of the room temperature XRD pattern with $\textit{P}\overline{1} $ space group is presented in Fig.\,\ref{fig:XRD}(a). The fitted neutron diffraction patterns at 295 K with both $\textit{P}2_{1}/\textit{n}$ and $\textit{P}\overline{1} $ space groups are presented in Fig.\,\ref{fig:XRD}(b).

Such a low-symmetry crystal structure corresponds to the $a^-b^-c^-$ octahedral tilt system in double perovskites \cite{20,21} and has been found previously in several related compounds, such as Ba$_{2}$LaRuO$_{6}$ \cite{22}, Ba$_{2}$NdMoO$_{6}$ \cite{23}, or Ba$_{2}$BiIrO$_{6}$ \cite{24}. The crystal structure consists of alternating octahedra of Cu and Ir in three crystallographic directions [Fig.\,\ref{fig:XRD}(c)]. The structure is derived from the distorted rock salt structure, which is one of the three common crystal structures of double perovskites, i.e. (i) random, (ii) rock salt, and (iii) layered \cite{19}. The relation between lattice parameters ($\sqrt{2}\textit{a}_{\rm p} \times 2\textit{a}_{\rm p} \times \sqrt{2}\textit{a}_{\rm p}$, where $\textit{a}_{\rm p}$ is the lattice parameter of the cubic perovskite structure) confirms the rock salt structure. However, in the present crystal structure, there are two different types of octahedra for both Cu (Cu1O$_{6}$ and Cu2O$_{6}$) and Ir (Ir1O$_{6}$ and Ir2O$_{6}$) sites with different amount of octahedral distortions and tiltings. We find that the  octahedra surrounding Cu atoms are more distorted than those around the Ir sites.  The two inequivalent octahedral sites of both Cu and Ir lead to two different types of alternating layers along the $\textit{b}$-axis. Among them, the first layer consists of Cu1O$_{6}$ and Ir1O$_{6}$ octahedra, and the second layer consists of structurally inequivalent Cu2O$_{6}$ and Ir2O$_{6}$ octahedra. A pictorial representation of this arrangement is displayed in Fig.\,\ref{fig:XRD}(c). The detailed crystallographic information as well as the bond-angle details (obtained from the XRD refinement) are listed in Table\,\ref{tab:latticeparameters} and \,\ref{tab:bondlengths}. An empirical measure for the distortion of such a perovskite structure is presented by the tolerance factor $\textit{t} = (\textit{d}_{A- O})/\sqrt{2}(\textit{d}_{B-O})$, where A and B are the two atoms of ABO$_{3}$ type perovskites \cite{25}. The ideal cubic perovskites correspond to $\textit{t} = 1$, and as the $\textit{t}$ value decreases further, the octahedra tilt and the crystal symmetry is reduced.  Generally, for $\textit{t}$ < 0.97, the crystal structure is expected to be monoclinic \cite{26}. For La$_{2}$CuIrO$_{6}$, $\textit{t}$ is calculated to be 0.8623. Interestingly, a similar value $\textit{t}$ = 0.8682 is also observed in the sister compound La$_{2}$CoIrO$_{6}$, whose crystal symmetry is found to be $\textit{P}2_{1}/\textit{n}$ \cite{27}. Though the ionic radius of Cu$^{2+}$ [0.73 \r{A}] is close to that of Co$^{2+}$ [0.745\,\AA], a longer $\langle$Ir -- O$\rangle$ bond length might play an influential role in the reduced $\textit{P}\overline{1}$ crystal symmetry of La$_{2}$CuIrO$_{6}$.

\begin{figure}[t]
\includegraphics[width=1\columnwidth]{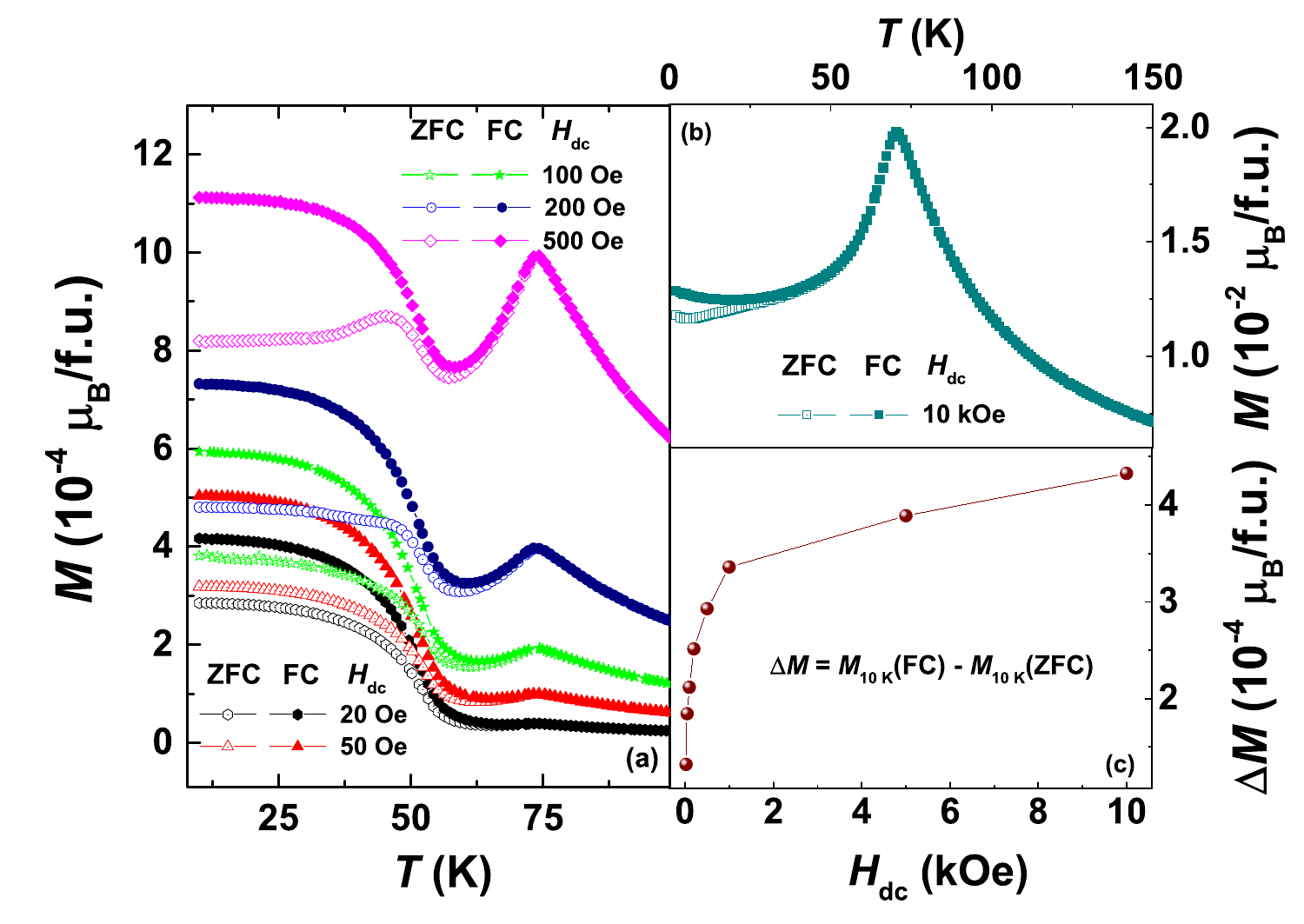}
\caption{(a) Temperature dependent field-cooled (FC) and zero-field-cooled (ZFC) magnetization of La$_{2}$CuIrO$_{6}$ at various applied \textit{dc} magnetic fields ($\textit{H}_{\rm dc}$) of 20, 50, 100, 200 and 500 Oe. The respective data for $\textit{H}_{\rm dc}$ = 10 kOe are presented in (b). Plot (c) shows the variation of the corresponding difference between the FC and ZFC magnetizations at 10 K with $\textit{H}_{\rm dc}$.}
\label{fig:dcmag}\vspace{-1pt}
\end{figure}

\begin{figure}[t]
\includegraphics[width=1\columnwidth]{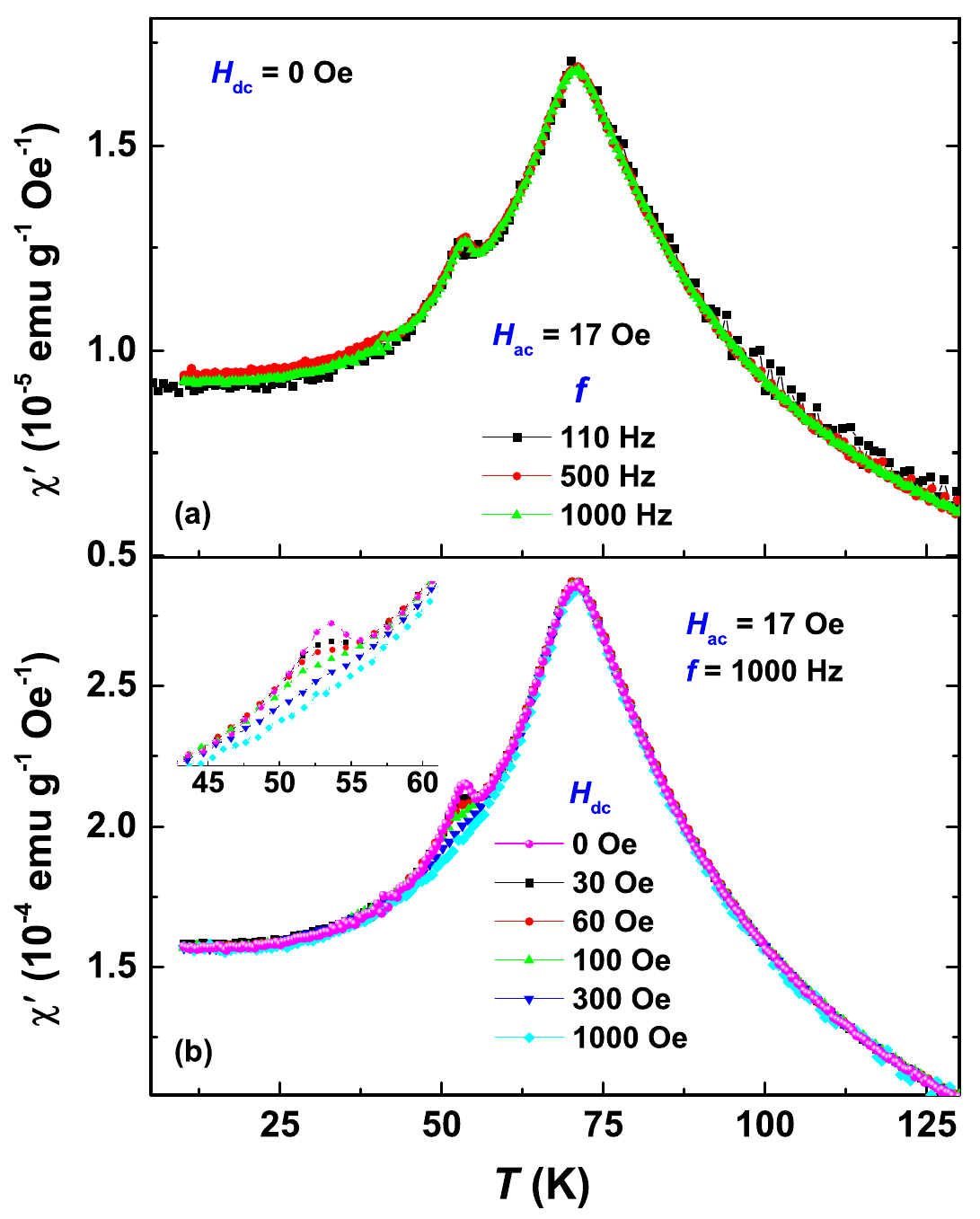}
\caption{(a) Temperature dependent real component of the \textit{ac} susceptibility [$\chi^{\prime}(\textit{T})$] of La$_{2}$CuIrO$_{6}$ with \textit{ac} field of $\textit{H}_{\rm ac}$ = 17 Oe and frequency, 110, 500 and 1000 Hz. (b) \textit{dc} biasing effect of $\chi^{\prime}(\textit{T})$ with a fixed \textit{ac} field. The corresponding inset show a zoomed-in view around $T_{2}$.}
\label{fig:acsus}\vspace{-10pt}
\end{figure}

\begin{figure}[b]
\includegraphics[width=1\columnwidth]{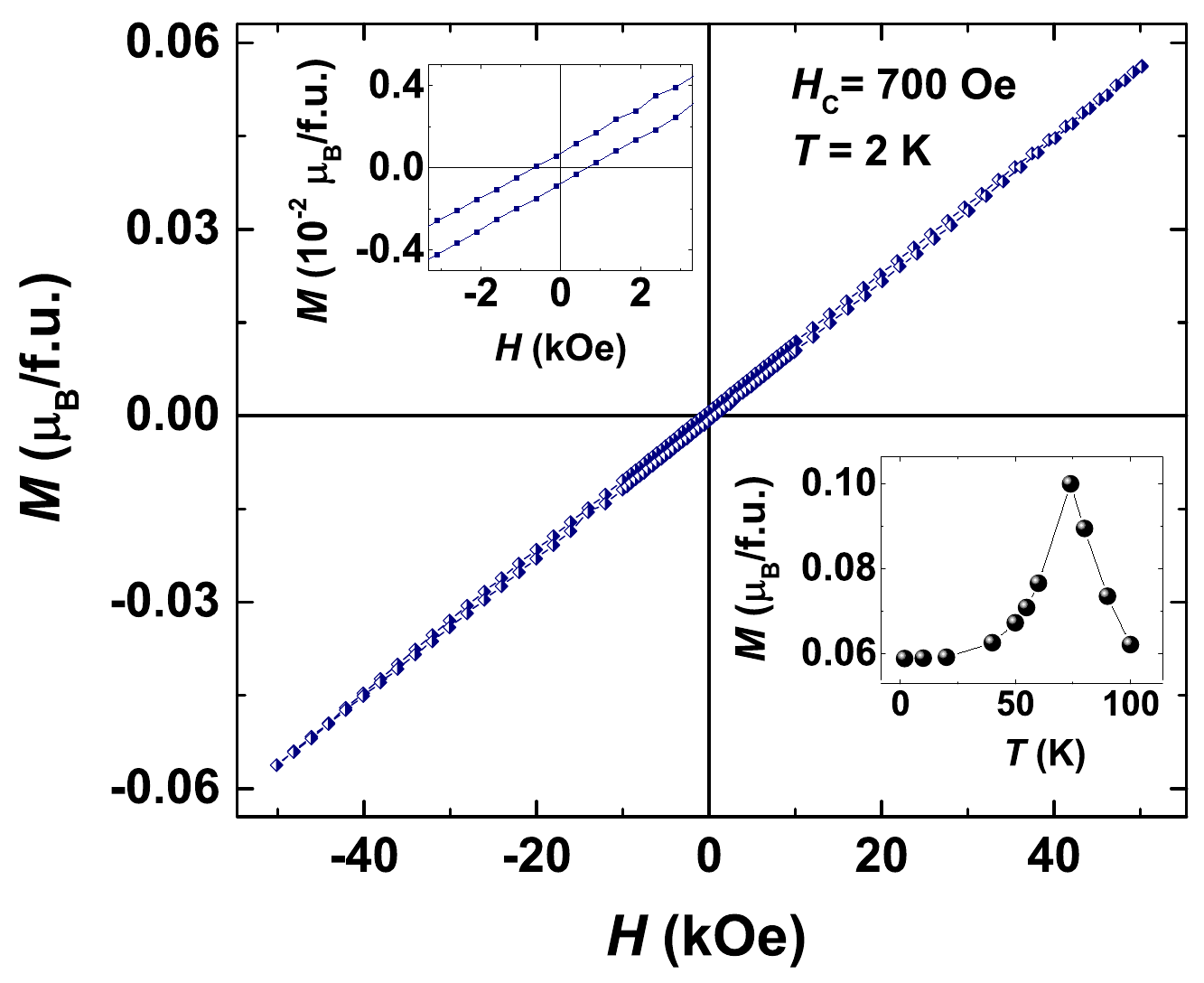}
\caption{$M - H$ curves of La$_{2}$CuIrO$_{6}$ at 2 K. The upper inset displays the corresponding zoomed in view at the low field region. The temperature dependence of the magnetization value at 5 T is presented in the lower inset.}
\label{fig:hys}\vspace{-5pt}
\end{figure}

\vspace{-10pt}\subsection{\textit{dc} Magnetization and \textit{ac} Susceptibility}\vspace{-1pt} 

Fig.\,\ref{fig:dcmag}(a) illustrates the temperature dependent field-cooled (FC) and zero-field-cooled (ZFC) magnetization [$M_{\rm FC}$ and $M_{\rm ZFC}$] for the La$_{2}$CuIrO$_{6}$ sample. As the sample is cooled from the high temperature paramagnetic state, a cusp typical for an antiferromagnetic phase transition appears around $T_{1} = 74$\,K. Upon further cooling, below $T_{2} = 54$\,K a weak ferromagnetic magnetization is observed in low external field. Two features are evident from the graph: (i) With the increase of the applied \textit{dc} magnetic field $H_{\rm dc}$, the $T_{1}$ transition becomes more intense. For $H_{\rm dc}$ = 10\,kOe [Fig.\,\ref{fig:dcmag}(b)], the $T_{2}$ feature is suppressed. (ii) The bifurcation of $M_{\rm FC}$ and $M_{\rm ZFC}$ below $T_{2}$ increases with increasing ${H}_{\rm dc}$. To realize whether the effect sustains at high field too, we plot the magnetization difference at 10 K, $\Delta M$ vs $H_{\rm dc}$ in Fig.\,\ref{fig:dcmag}(c), where $\Delta M$ = $M_{\rm 10 K}$(FC)-- $M_{\rm 10 K}$(ZFC). It is clear that $\Delta M$ increases rapidly for small fields $\leq$ 1 kOe and shows a saturation tendency at larger fields. 

Figures \ref{fig:acsus}(a) and (b) present the real part of the temperature dependent \textit{ac} susceptibility $\chi^{\prime}(\textit{T})$ for the La$_{2}$CuIrO$_{6}$ sample measured in various conditions. Here, the $T_{1}$ transition appears as a large peak, and upon cooling, $T_{2}$ develops as a small shoulder. Fig.\,\ref{fig:acsus}(a) displays the frequency dependence of $\chi^{\prime}(\textit{T})$ with a fixed applied \textit{ac} field of $\textit{H}_{\rm ac}$ = 17 Oe. It is clear that all the $\chi^{\prime}(\textit{T})$ curves lie on top of each other and neither $T_{1}$ nor $T_{2}$ peaks exhibit any frequency dependence. This rules out that either $T_{1}$ or $T_{2}$ is associated with any kind of conventional glassy transition \cite{28,29}. Fig.\,\ref{fig:acsus}(b) illustrates the effect of \textit{dc} biasing field on $\chi^{\prime}(\textit{T})$. Here a persistent \textit{dc} biasing field ($\textit{H}_{\rm dc}$ = 0, 30, 60, 100, 300 and 1000 Oe) is superimposed with an \textit{ac} magnetic field of $\textit{H}_{\rm ac}$ = 17 Oe, and $\chi^{\prime}(T)$ was recorded in ZFC mode with a frequency $f = 1000$ Hz. It is evident that $T_{1}$ is not affected by the applied \textit{dc} biasing field, a characteristic feature of an AFM transition. However, the increasing \textit{dc} biasing field damps the $T_{2}$ peak intensity.

\begin{figure}[t]
\includegraphics[width=1\columnwidth]{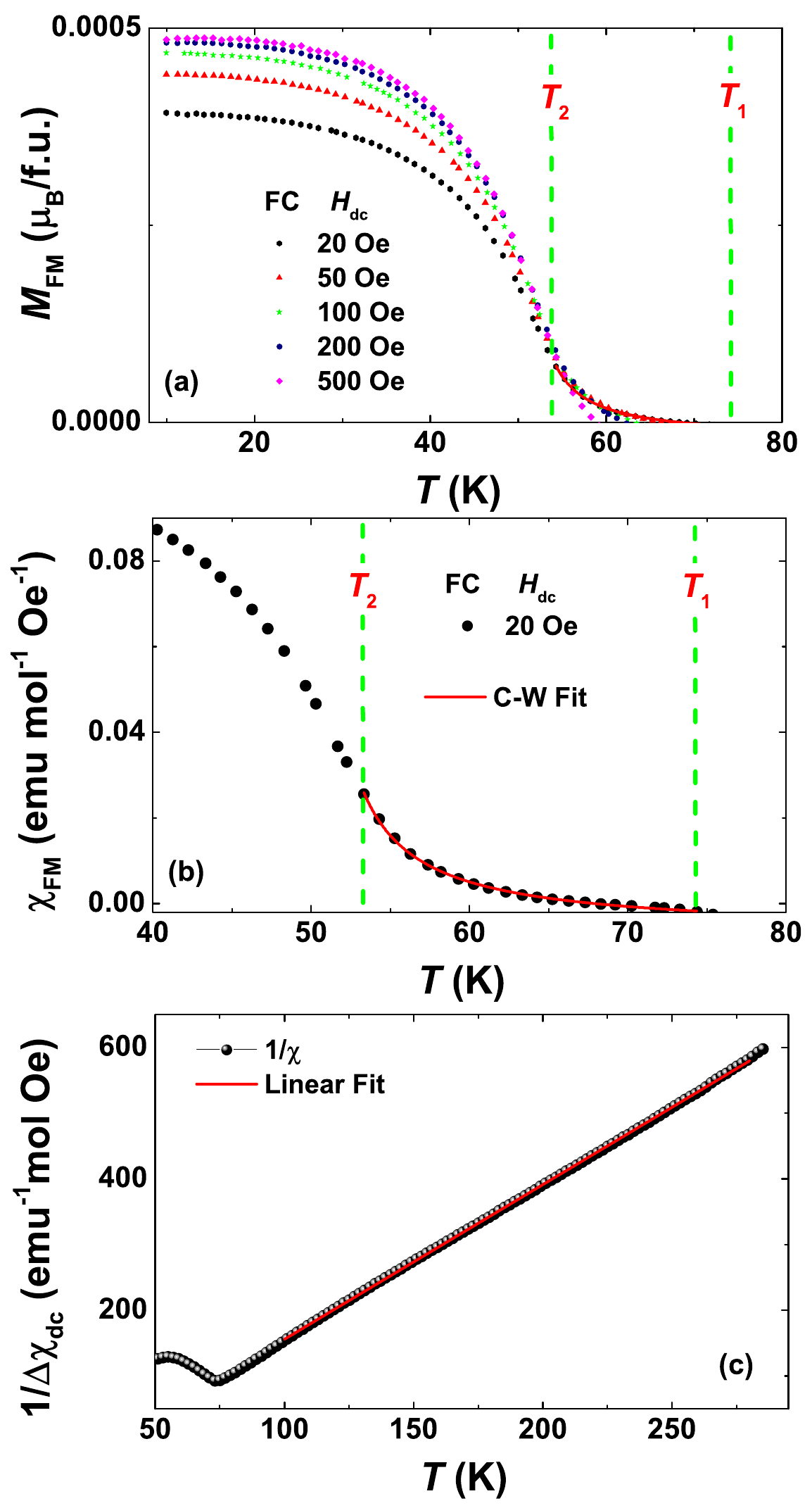}
\caption{(a) The deduced ferromagnetic magnetization component in La$_{2}$CuIrO$_{6}$ for different magnetic fields between 20 and 500 Oe \cite{EndNote}. (b) The magnetic susceptibility for $H_{\rm dc} = 20$\,Oe around the two transitions. The red line is a fit to a Curie-Weiss [C-W] law (for details see text). (c) Temperature dependent reciprocal \textit{dc} susceptibility with straight line as the Curie-Weiss fit. Here $\Delta\chi_{\rm dc} =\chi_{\rm dc}- \chi_{0}$, where $\chi_{0}$ is a temperature independent contribution to $\chi_{\rm dc}$.}
\label{fig:FMcomp}\vspace{-5pt}
\end{figure}

Figure \ref{fig:hys} shows the $M - H$ hysteresis loop of La$_{2}$CuIrO$_{6}$ at 2 K. The magnetization linearly increases with the applied magnetic field as is usually seen in other AFM materials. Interestingly, the system possesses a loop opening with the critical field, $\textit{H}_{\rm C} \sim 700$ Oe as evident from the upper inset of Fig.\,\ref{fig:hys}. This indicates the presence of domains with weak FM correlations in La$_{2}$CuIrO$_{6}$ along with the dominating AFM order. The temperature dependence of the magnetization value at 50 kOe [the lower inset of Fig.\,\ref{fig:hys}] proves the dominance of the AFM ordering under this field. We find no anomaly around $T_{2}$, but the curve peaks around $T_{1}$. 

In order to extract the FM magnetization component below $T_{1}$ from the low field magnetization data, the AFM component was modeled using the normalized high field (10~kOe) magnetization data shown in Fig.\,\ref{fig:dcmag}(b) for $2\,{\rm K} \leq T \leq T_{1}$. As seen from Fig.\,\ref{fig:FMcomp}(a), the FM component starts to saturate with absolute values of the order of $5 \times 10^{-4}\mu_{\rm B}$/f.u. for temperatures $T < \text{T}_{2}$. This order of magnitude points towards a small canting of the AFM ordered moments of the Cu and Ir sublattices (see below). Fig.\,\ref{fig:FMcomp}(b) clearly shows that $M_{\rm FM}$ follows a Curie-Weiss law for $T_1 \leq T \leq T_2$. A similar behavior has been found for other AFM systems, where the Dzyaloshinsky-Moriya interaction results in a spin canting and thus a FM magnetization component following a Curie-Weiss law at low temperatures \cite{31}. 

The paramagnetic phase above $T_{1}$ is further analyzed by plotting the temperature dependent inverse susceptibility $[\chi_{\rm dc}(\textit{T})]^{-1}$ derived using [$\textit{M}_{\rm ZFC}/\textit{H}_{\rm dc}]^{-1}$. In Fig.\,\ref{fig:FMcomp}(c) we present $\Delta\chi_{\rm dc}^{-1}$ vs $T$ where $\Delta\chi_{\rm dc}$ is computed by subtracting a temperature independent background contribution from $\chi_{\rm dc}$. The high temperature data, in the $100\,\text{K} < \textit{T} < 280\,\text{K}$ window, nicely fit with the Curie-Weiss law yielding an effective magnetic moment of $\mu_{\rm eff}$ = 1.84 $\mu_{\rm B}$/f.u. Considering the spin configuration of Cu$^{2+}$ -- Ir$^{4+}$, as reported earlier \cite{32}, the effective moment $\mu_{\rm eff}$ is found to be less than  2.83 $\mu_{\rm B}$/f.u. for an assumed $\textit{J}$ = 1 state. Interestingly the Curie-Weiss temperature is found to be $\theta_{\rm CW}$ = + 33.7 K irrespective of the dominating AFM interaction, as the signal from FM component dominates in \textit{dc} magnetization data. A similar kind of double transition behavior is also observed for La$_{2}$ZnIrO$_{6}$\cite{33}, where the spin structure was found to be of a canted AFM type with different spin canting angles at the two transitions.

\vspace{-5pt}\subsection{Neutron Diffraction}\vspace{-5pt}

\begin{figure}[t]
\includegraphics[width=1\columnwidth]{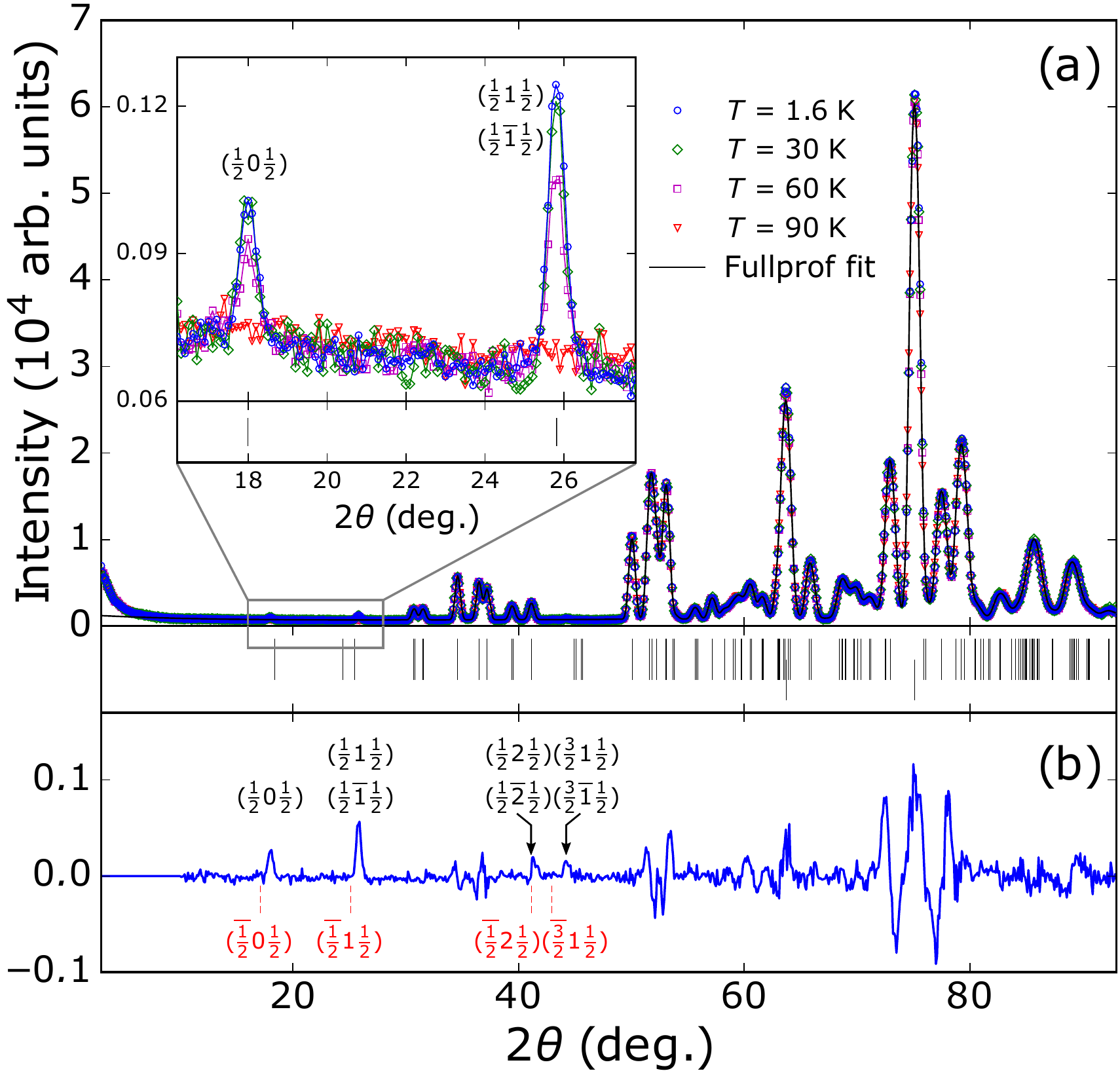}
\caption{(a) Neutron powder diffraction patterns of La$_{2}$CuIrO$_{6}$ as a function of the scattering angle (2$\theta$) for different temperatures. The black solid line is the best Rietveld fit to the triclinic $\textit{P}\overline{1} $ space group of the 90 K reference dataset measured above $T_{1}$. Calculated structural peak positions are shown with vertical lines below the plot. The two strongest reflections arise from the aluminum of the sample container (their corresponding marks are shifted downwards). The inset shows a blown-up fragment of the diffraction pattern at low scattering angles, where temperature-dependent magnetic Bragg peaks can be seen around 2$\theta$ = 18.0$^\circ$ and 25.8$^\circ$. They can be assigned to the commensurate $\bigl(\frac{1}{2}\, 0\, \frac{1}{2}\bigr)$ and $\bigl(\frac{1}{2}\, \pm\!1\, \frac{1}{2}\bigr)$ reflections whose calculated positions are shown below with two vertical lines. (b) The difference of the low- and high-temperature datasets, from which the respective difference of the structural models has been subtracted to eliminate the effect of thermal expansion. Two additional magnetic Bragg peaks marked by arrows can be seen. Positions of the forbidden magnetic reflections as described in the text are shown below the curve with dashed lines.}
\label{fig:Neutron}\vspace{-5pt}
\end{figure}

In order to verify the nature of the crystal structure and to understand the microscopic nature of the magnetic ordering of La$_{2}$CuIrO$_{6}$, we performed temperature dependent neutron diffraction measurements. Fig.\,\ref{fig:Neutron}(a) presents the diffraction data, collected with an incident neutron wavelength of 2.458 \r{A} at four different temperatures above and below the magnetic transitions. The temperature dependent study reveals no structural transition over the entire temperature range.

The spin configuration of La$_{2}$CuIrO$_{6}$ is determined by analyzing the low-temperature neutron diffraction data. At temperatures below $T_{1}$, we observe the appearance of additional commensurate magnetic reflections, which are best seen at low scattering angles [Fig.\,\ref{fig:Neutron}(a), inset]. They can be indexed as $\bigl(\frac{1}{2}\, 0\, \frac{1}{2}\bigr)$ and $\bigl(\frac{1}{2}\, \pm\!1\, \frac{1}{2}\bigr)$, suggesting that the magnetic structure in La$_{2}$CuIrO$_{6}$ is antiferromagnetic with a propagation vector $\bigl(\frac{1}{2}\, 0\, \frac{1}{2}\bigr)$, similar to the one previously reported for the Co(II) sublattice in LaSrCoIrO$_{6}$ \cite{27} or Sr$_{2}$CoOsO$_{6}$ \cite{34}. In addition, much weaker higher-order magnetic Bragg peaks  $\bigl(\frac{1}{2}\, \pm\!2\, \frac{1}{2}\bigr)$ and  $\bigl(\frac{3}{2}\, \pm\!1\, \frac{1}{2}\bigr)$ can be seen in Fig.\,\ref{fig:Neutron}(b), where we present the difference of the 1.6 and 90\,K datasets. Since the contribution from nuclear scattering cannot be fully eliminated in such a subtraction due to the thermal expansion of the sample, here we have additionally subtracted the difference of the corresponding structural models in order to minimize the contamination to the magnetic signal. The absence of magnetic reflections at  $\bigl(\overline{\frac{1}{2}}\, 0\, \frac{1}{2}\bigr)$ (2$\theta \approx$ 17.1$^\circ$) and $\bigl(\overline{\frac{1}{2}}\,  \pm\!1\, \frac{1}{2}\bigr)$ (2$\theta \approx 25.2^{\circ}$), as denoted at the bottom of Fig.\,\ref{fig:Neutron}(b) by the dashed lines, indicates that the magnetic structure must be non-collinear. 

A satisfactory description of the magnetic diffraction pattern could be obtained by assuming an orthomagnetic spin structure with collinear AFM spin arrangement in every \textit{ac}-- plane (Cu1Ir1 or Cu2Ir2) and mutually orthogonal spin orientations in neighboring planes, as shown in Fig.\,\ref{fig:XRD}(c). However, due to the large number of structural reflections resulting from the low crystal symmetry and, as a consequence, a small number of clearly observable magnetic Bragg peaks, alternative solutions with a lower magnetic symmetry cannot be excluded. In our model, the fitted values of AFM ordered moments from the Rietveld refinement are 0.69 and 0.45 $\mu_{\rm B}$ on the Cu and Ir sites, respectively.  The model also allows for a weak FM contribution, as the resulting magnetic Bragg intensity would fall on top of the structural Bragg reflections and would be therefore unobservable in the present unpolarized neutron diffraction experiment. A systematic polarized-neutron study or measurement on a La$_{2}$CuIrO$_{6}$ single crystal might be helpful to get a more complete picture of the magnetic structure. The fact that magnetic Bragg peaks are present already at $\textit{T} = 60$\,K and do not change qualitatively upon cooling down to the base temperature indicates that the AFM structure sets in at the higher-temperature transition ($T_{1} \approx 74$\,K). No visible change in the neutron diffraction patterns on the second anomaly at $T_{2} \approx  54$\,K, as observed in the \textit{ac} and \textit{dc} susceptibility measurements, suggests a weak change in the magnetic intensity well below the detection limit.

\vspace{-5pt}\subsection{Specific Heat}\vspace{-5pt}

\begin{figure}[b]
\includegraphics[width=1\columnwidth]{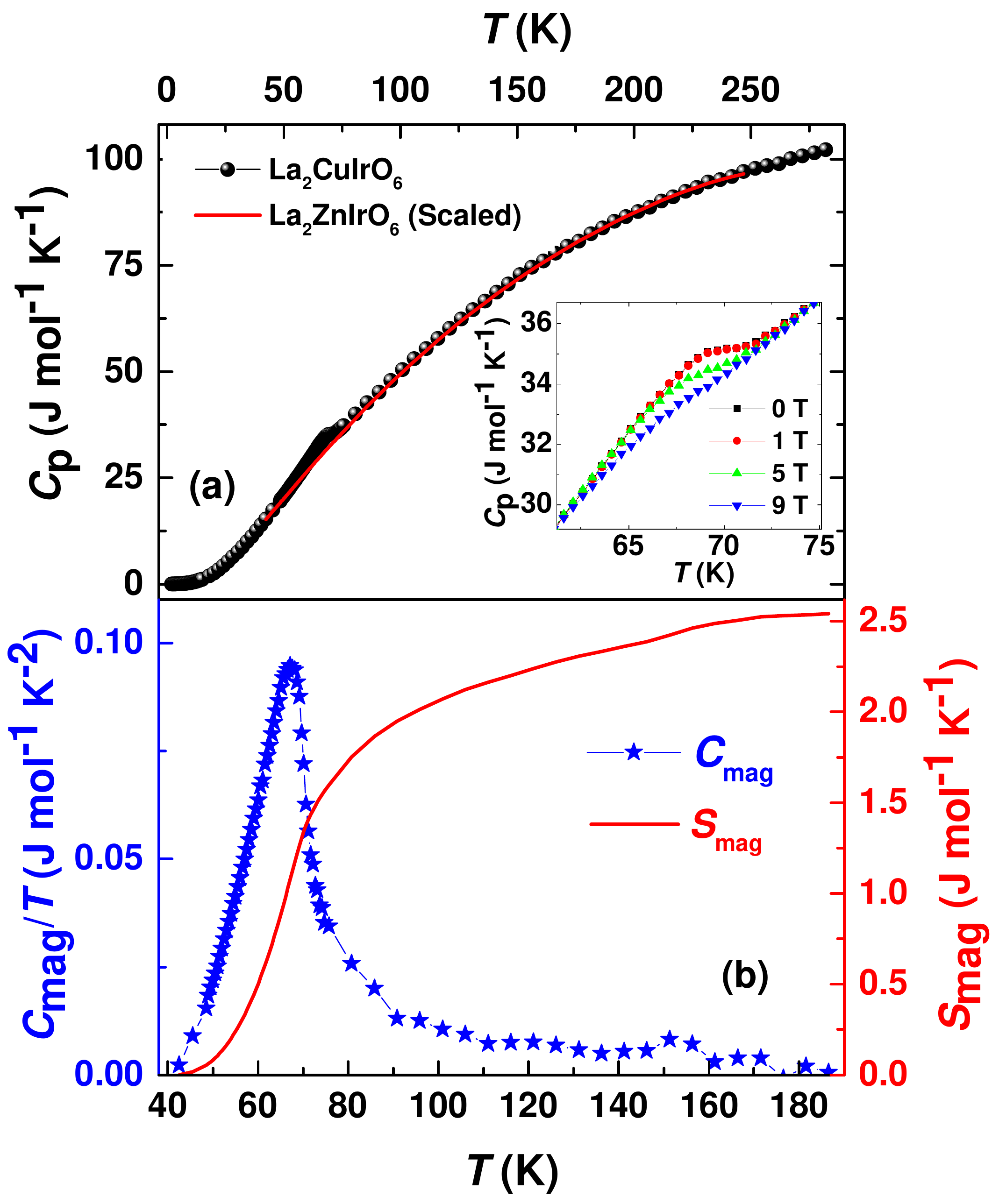}
\caption{(a) Temperature dependent specific heat [$C_{P}(T)$] of La$_{2}$CuIrO$_{6}$ at zero applied magnetic field (black symbol) and the respective one of La$_{2}$ZnIrO$_{6}$ for $T > 40$ K (red line). The latter was scaled by a factor of 1.017 (see text) in order to estimate the phononic contribution of La$_{2}$CuIrO$_{6}$. Inset shows the effect of external magnetic field on C$_{P}(\textit{T}$). (b) Zero-field magnetic specific heat of La$_{2}$CuIrO$_{6}$ plotted as $C_{\rm mag}/T$ vs $T$ (left scale) and the deduced temperature dependent magnetic entropy [$S_{\rm mag}(T)$] (right scale).}
\label{fig:Cp}\vspace{-5pt}
\end{figure}

The magnetic behavior of La$_{2}$CuIrO$_{6}$ is further analyzed by the temperature dependent specific heat [$C_{P}(T)$] measurement. Fig.\,\ref{fig:Cp}(a) presents the corresponding $C_{P}(T)$ measured in zero magnetic field. Two features are discernible from the figure, (i) $C_{P}(\textit{T})$ shows no anomaly around $T_{2}$, and (ii) a hump-like behavior appears around $T_{1}$. External magnetic field up to 9 T broaden and gradually shift the anomaly to lower temperatures [inset of Fig.\,\ref{fig:Cp}(a)], similar to what has been observed in other perovskite oxides \cite{35,36}.

To understand the magnetic interactions in La$_{2}$CuIrO$_{6}$, the high temperature specific heat data are analyzed by plotting the associated excess specific heat, $C_{\rm mag}(T)$ in Fig.\,\ref{fig:Cp}(b). Here we subtract the lattice contribution from the experimental $C_{P}(T)$ data by measuring $C_{P}(T)$ of the structural analog compound La$_{2}$ZnIrO$_{6}$, which undergoes magnetic transitions at 7.3 and 8.5 K \cite{33}, well below those of La$_{2}$CuIrO$_{6}$. Due to the additional (field-dependent) short-range correlations, which are present up to 40 K, the $C_{P}(T)$ of La$_{2}$ZnIrO$_{6}$ was used as an approximation for the lattice contribution in the temperature range of $40\,\text{K} \leq T \leq 250$\,K only. Since the mass and volume of the Zn and Cu compounds are different, the experimental La$_{2}$ZnIrO$_{6}$ specific heat curve must be scaled. An effective Debye temperature can be expressed as \cite{37,38}:
\begin{equation}
\theta_{D}=L\Big(\frac{T_{\rm m}}{MV^{2/3}}\Big)^{1/2},
\label{Eq:Cp1}
\end{equation}
where $L $is the Lindemann constant, $T_{\rm m}$ the melting temperature, $M$ the atomic weight, and $V$ the atomic volume. If $L$ and $T_{\rm m}$ are assumed nearly the same for both compounds, we get the relationship:
\begin{equation}
\Big(\dfrac{\theta_{\rm D1}}{\theta_{\rm D2}}\Big)^{2}=\dfrac{M_{2}V_{2}^{2/3}}{M_{1}V_{1}^{2/3}}.
\label{Eq:Cp2}
\end{equation}
By applying this relation, we obtain that $\theta_\text{D}^{\text{La}_{2}\text{CuIrO}_{6}} = 1.017 \times \theta_\text{D}^{\text{La}_{2}\text{ZnIrO}_{6}}$. The scaled La$_{2}$ZnIrO$_{6}$ data are shown as a red line in Fig. 7(a). It is clear that the curves nicely overlap each other in the high-temperature region, except for the La$_{2}$CuIrO$_{6}$ magnetic transition region.

After subtracting the scaled La$_{2}$ZnIrO$_{6}$ data from La$_{2}$CuIrO$_{6}$, we obtain the magnetic contribution to the specific heat, which is plotted as $C_\text{mag}/T$ as a function of temperature in Fig.\,\ref{fig:Cp}(b) (left scale). A sharp peak is observed around $T \approx 67$\,K. Using an entropy-balance construction, a transition temperature $T_{\rm N} \approx 74$\,K is extracted. The magnetic entropy ($S_{\rm mag}$) for La$_{2}$CuIrO$_{6}$ was evaluated by integrating the zero-field specific heat data, as shown in Fig.\,\ref{fig:Cp}(b) (right scale). The maximum value of $S_{\rm mag}$ is calculated as 2.54~J\,mol$^{-1} \text{K}^{-1}$, which is much less than the full spin entropy of $R\,\text{ln}(2J+1) \approx 9.134$~J\,mol$^{-1} \text{K}^{-1}$. Here $J$ is the total angular momentum. Interestingly, $C_{\rm mag}(T)$ retains small positive values up to $T \approx 120-160$ K, which points to short-range magnetic correlations appearing at much higher temperatures than $T_{1}$ [see Fig.\,\ref{fig:Cp}(b)]. These short-range correlations must reduce the entropy associated with the AFM transition, however, other effects such as quantum fluctuations or partial spin alignment might additionally play a role. A similar scenario was observed by Lees \textit{et al.} \cite{39} in Pr$_{0.6}$(Ca$_{1-x}$Sr$_{x})_{0.4}$MnO$_{3}$ compounds where the short-range magnetic correlations due to Mn cluster or formation of Jahn-Teller polarons at higher temperature reduces the magnetic entropy associated with the FM transition. The clear ``cutoff--temperature'' for these short-range correlations in La$_{2}$CuIrO$_{6}$ could be slightly smaller due to limits within the phononic contribution, which was calculated for $T > 40$ K. As will be discussed in the next section, this is in agreement with the ESR results.

\begin{figure}[b]
\includegraphics[width=1\columnwidth]{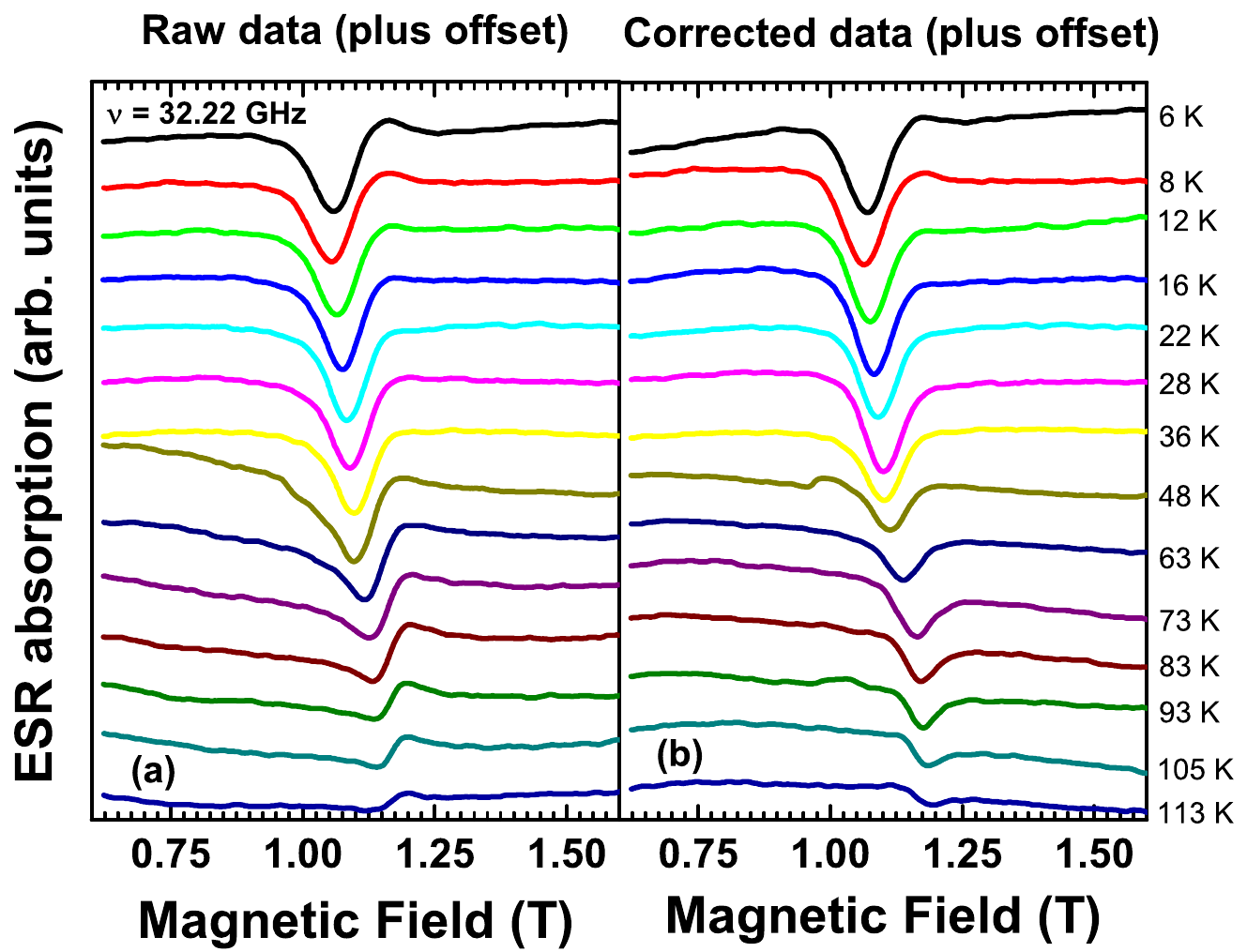}
\caption{Selected ESR spectra of La$_{2}$CuIrO$_{6}$ at different temperatures at a fixed frequency of $\nu$ = 32.22 GHz [(a) as measured spectra; (b) after correction by the analyzer software]. The spectra are shifted vertically for clarity.}
\label{fig:ESR1}\vspace{-5pt}
\end{figure}

\vspace{-5pt}\subsection{ESR Experiment}\vspace{-5pt}

We performed high-field multi-frequency ESR spectroscopy to obtain insights on the dynamics of the coupled Cu and Ir spins in La$_{2}$CuIrO$_{6}$. The spectra were recorded at a given fixed frequency, $\nu$, while sweeping the magnetic field up and down. Fig. 10 presents the temperature variation of the characteristic ``as  measured'' signals at $\nu = 32.22$~GHz. The ESR lines were fitted with the function \cite{40}:
\begin{multline}\label{Eq:ESR}
F(H)=\dfrac{A\Delta H}{[4(H-H_{0})^{2}+\Delta H^{2}]\pi}.\cos(\alpha)\\+
\dfrac{A(H-H_{0})}{[4(H-H_{0})^{2}+\Delta H^{2}]\pi}.\sin(\alpha)+BH+C.
\end{multline}
Here the first two terms are the absorption and dispersion parts of the complex Lorentzian function, respectively, with the amplitude A, the width $\Delta H$, and the resonance field $\textit{H}_{0}$. The last two terms with fit parameters $\textit{B}$ and $\textit{C}$ account for an instrumental linear in-field non-resonant background and a constant offset, respectively. The parameter $\alpha$ defines the relative contributions of absorption and dispersion in the total signal. A mixing of the real and imaginary parts with the increase in temperature is visible in the spectra in Fig.\,\ref{fig:ESR1}. It arises due to a complex impedance of the broadband wave-guides, which can depend on temperature due to thermal expansion. Since the network analyzer measures both the amplitude and the phase shift of the signal, such distortion of the line shape can be corrected with the analyzer's software [see Fig. \,\ref{fig:ESR1}(b)]. 

\begin{figure}[t]
\includegraphics[width=1\columnwidth]{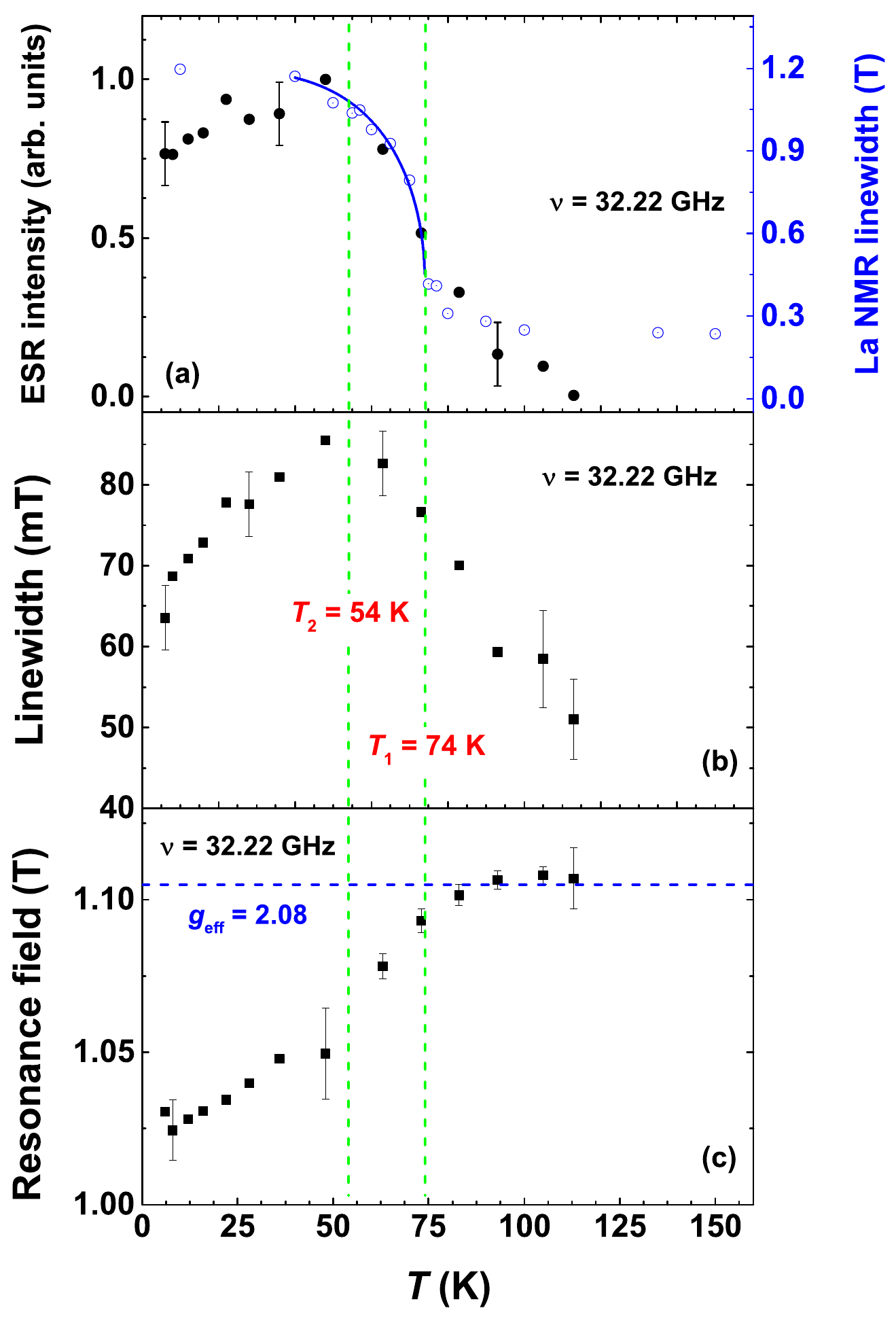}
\caption{Temperature dependence of the integrated intensity $\textit{I} \sim \textit{A}\Delta \textit{H}$ (a), linewidth (b) and resonance field (c) determined from the ESR spectra at $\nu$ = 32.22 GHz in Fig.\,\ref{fig:ESR1} using the fit function (\ref{Eq:ESR}). In (a) the temperature dependence of the $^{139}$La NMR linewidth (open circles) and the calculated dependence of the order parameter (solid line) are shown for comparison (see section ``NMR Measurements'').}
\label{fig:ESR2}\vspace{-5pt}
\end{figure}

The results of the fit are summarized in Fig.\,\ref{fig:ESR2}. Several aspects of the temperature evolution of the ESR parameters are noticeable. The ESR signal becomes first visible at a temperature of $\sim$ 113 K which is substantially larger than the AFM ordering temperature $T_1 = T_{\rm N} = 74$\,K. The integrated intensity of the signal determined for the Lorentzian line as $\textit{I}$ = 1.57 A$\Delta H$ \cite{40} smoothly increases and reaches a saturation around $T_2 = 54$\,K. The $\Delta H(\textit{T})$ dependence exhibits a dome-like shape with a broad maximum centered around $T_{2}$. The resonance field $\textit{H}_{0}$ stays practically constant down to $\sim$ 90 K corresponding to an effective $g$-factor $g_{\rm eff}$ = $h\nu/\mu_{\rm B}\textit{H}_{0} \approx 2.08$. Below $T_{1}$ the $\textit{H}_{0}$ gradually shifts to lower values. 

\begin{figure}[b]
\includegraphics[width=1\columnwidth]{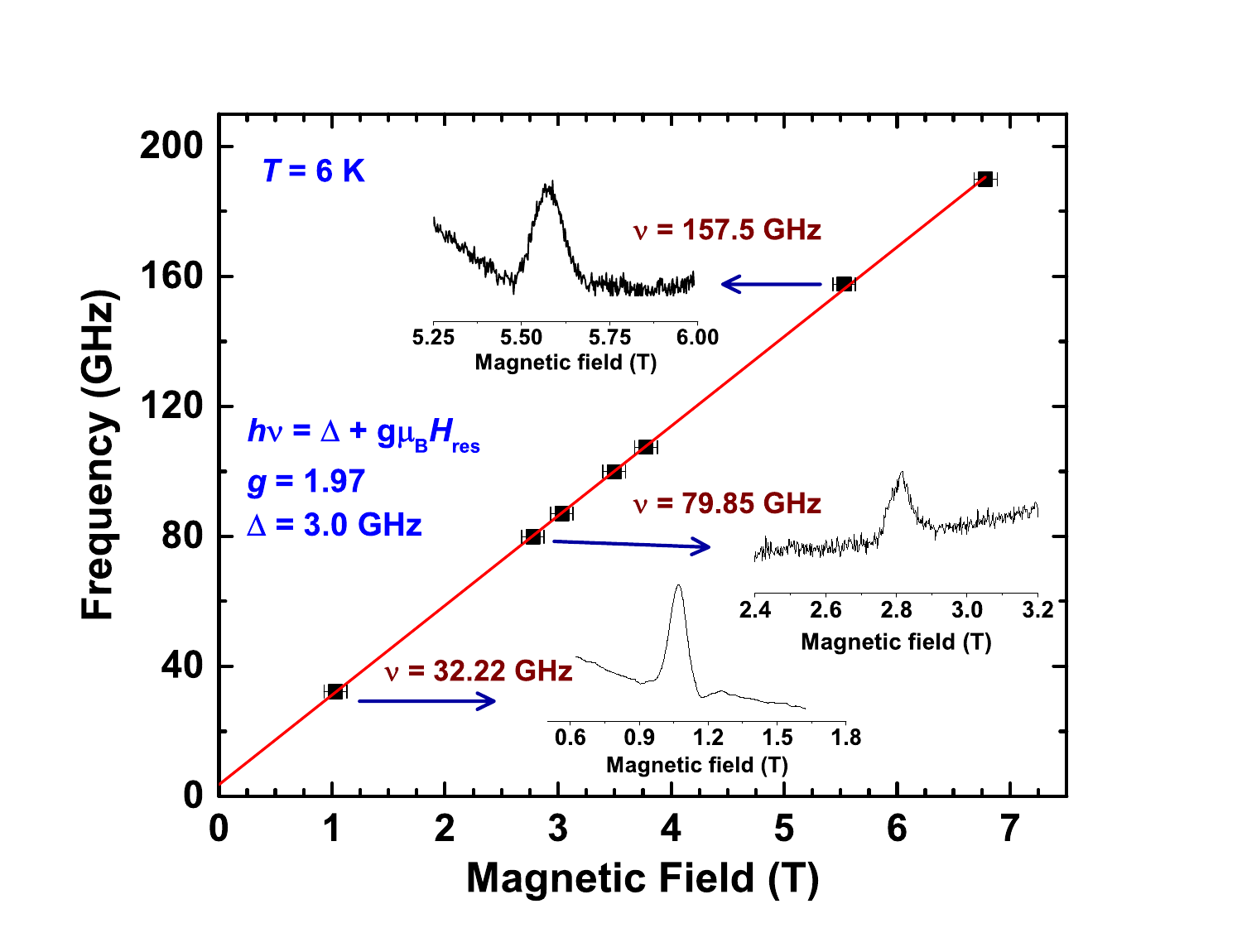}
\caption{Frequency $\nu$ vs $\textit{H}_{0}$ dependence of the ESR signal at $T = 6$\,K (symbols) and a fit to the function $h\nu = \Delta + g\mu_{\rm B}\textit{H}_{0}$ (solid line). Representative ESR signals (absorption part) are shown in insets.}
\label{fig:ESR3}\vspace{-5pt}
\end{figure}

\begin{figure}[b]
\includegraphics[width=1\columnwidth]{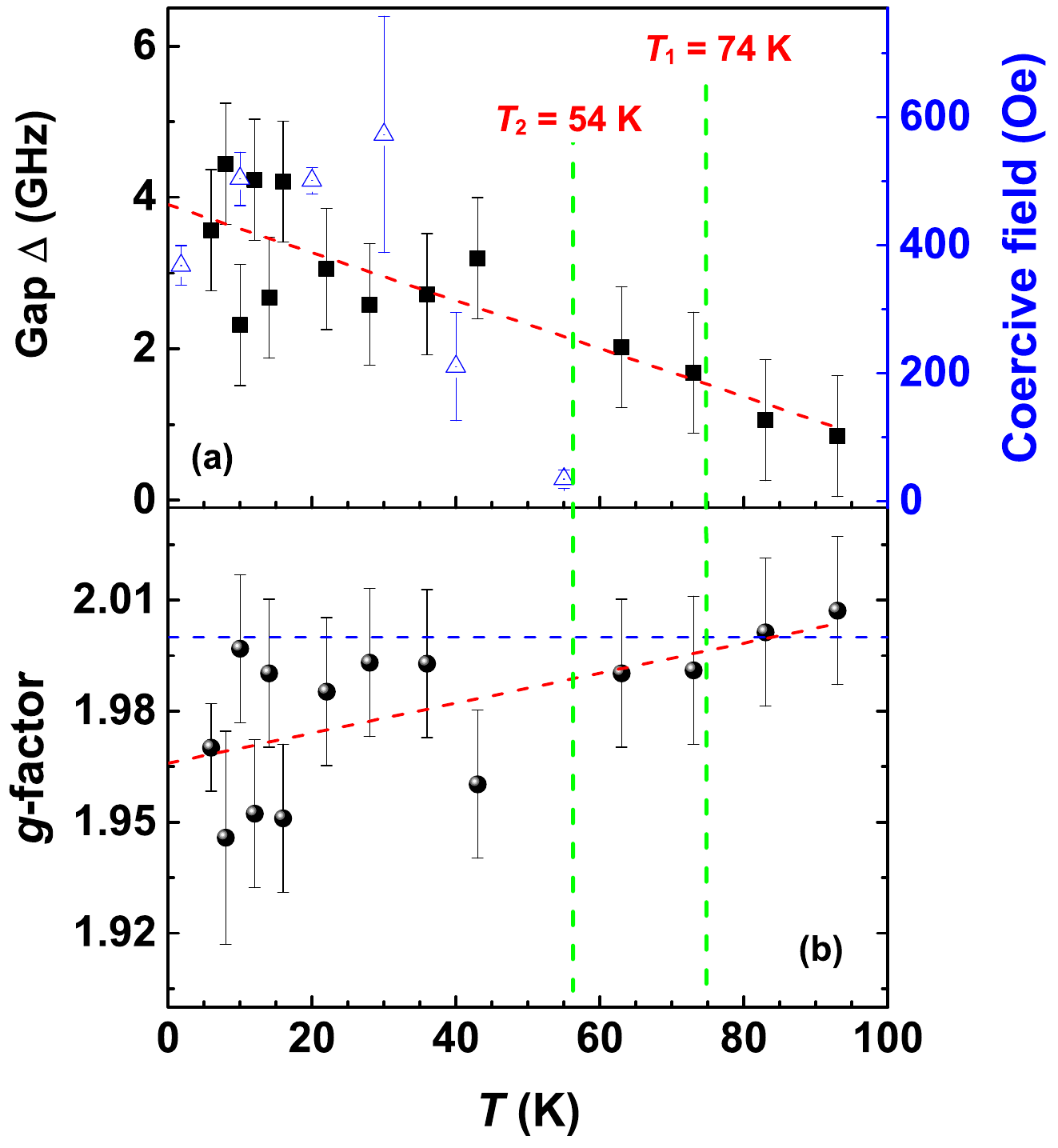}
\caption{(a) Temperature dependence of the ESR energy gap $\Delta$ and (b) the $\textit{g}$-factor obtained from the $\nu(\textit{H}_{0})$ dependencies. Dashed red lines are guides for the eye. Horizontal dashed line in (b) panel denotes the spin-only $\textit{g}$-factor of 2. The temperature dependence of the coercive field $\textit{H}_{C}$ is plotted as open triangle in (a).}
\label{fig:ESR4}\vspace{-5pt}
\end{figure}

Further important information on the $g$-factor and a possible frequency (energy) gap for ESR excitations can be determined from ESR measurements at several frequencies performed at a given constant temperature. A representative frequency $\nu$ vs $\textit{H}_{0}$ dependence (resonance branch) taken at 6 K is shown in Fig.\,\ref{fig:ESR3}. Experimental data points clearly follow a linear dependence which can be parameterized as $h\nu$ = $\Delta + g\mu_{\rm B}\textit{H}_{0}$. The slope of the curve is determined by the $\textit{g}$-factor, and the offset $\Delta$, where the fit function meets the frequency axis at $\textit{H}_{0}$ = 0, defines the minimal energy for the ESR excitation. The fit gives $g$ = 1.97 and $\Delta$ = 3.0 GHz. A similar fitting procedure applied to the $\nu(\textit{H}_{0})$ data-- sets collected at different temperatures yields the $\textit{T}$-dependence of $\textit{g}$ and $\Delta$, as shown in Fig.\,\ref{fig:ESR4}.

In the high temperature regime for $T \gg T_1$ no ESR signal can be detected, suggesting that Ir and Cu spins relax fast most likely due to a substantial anisotropy of the spin-spin interactions. In the case of Ir pseudospins, this happens due to a strong coupling with phonons \cite{41}. The fact that the ESR signal appears below $\sim$ 113 K, above $T_{1}$, signifies a crossover from a purely paramagnetic resonance regime to the development of a collective resonance mode of correlated spins and pseudospins in the quasi-static (on the fast ESR time scale) short-range ordered state of La$_{2}$CuIrO$_{6}$. This also explains why the magnetic entropy is so low at $T_{1}$ in $C_{P}(T)$, though indications of any short-range order are not visible in the CW fitting of the \textit{dc} susceptibility over 100 -- 280 K. In contrast to paramagnets, where spins relax individually, the collective excitation mode, i.e. the precession of the total magnetization of the spin system, is more difficult to broaden by the above-mentioned relaxation mechanisms. An opening of the gap for the ESR excitations [Fig.\,\ref{fig:ESR4}(a)] gives evidence that the resonating spin system is static on the time scale of ESR of the order $\sim$\,10 ns and suggests an easy-axis type of magnetic anisotropy \cite{42}. The smallness of the gap, which at the lowest temperature amounts to at most $\sim$\,4 GHz (0.2 K or 0.016 meV) suggests that it might not be necessarily an AFM spin wave gap at the magnetic zone center, i.e. a magnetic anisotropy gap for the uniform oscillation of the AFM sublattices. The value of such an AFM gap for the layered cuprates and iridates is at least a few meV (see e.g. \cite{43,44,45}) or even substantially larger \cite{46}. Certainly a situation in La$_{2}$CuIrO$_{6}$ might be different due to a different crystal structure comprising two interpenetrating Cu and Ir perovskite sublattices. Specifically, in the related double perovskites La$_{2}$MgIrO$_{6}$ and La$_{2}$ZnIrO$_{6}$ a gap for spin excitations was reported amounting to 2.6 meV and 2.1 meV, respectively \cite{47}. The observed ESR signal gives evidence for spin excitations with much smaller energy in La$_{2}$CuIrO$_{6}$. It is reasonable to attribute this mode to a resonance of the uncompensated transverse net moments in a non-collinear antiferromagnet, which is, in fact, just another oscillation mode of coupled non-collinear AFM sublattices \cite{42}. Unlike the AFM anisotropy gap which is given by the anisotropic part of the superexchange interaction between the spins, the energy gap for oscillations of the net moments is determined by smaller residual anisotropies and often lies in the GHz frequency range (see., e.g., \cite{48,49}). The fact that the observed resonance mode occurs in the short-range ordered regime above $T_{1} = 74$\,K and continuously develops upon lowering the temperature suggests that uncompensated ``weak ferromagnetism'', i.e. a small noncollinearity of the AFM sublattices, is inherent to the AFM state of La$_{2}$CuIrO$_{6}$. The temperature $T_{2} = 54$\,K identified in the magnetometry measurements can be presumably considered as a kind of ``blocking'' temperature for the net moments at which the magnetic anisotropy wins against the entropy effect of temperature. It shows up in a maximum of the $\Delta H(T)$ dependence (correlation peak) and saturation of the intensity of the ESR signal (Fig.\,\ref{fig:ESR2}). Below this blocking temperature, the magnetization curve $M(H)$ starts exhibiting a small hysteresis [Fig.\,\ref{fig:hys}]. The $\textit{T}$-dependence of the respective coercive field $\textit{H}_{\rm C}$ is plotted on Fig.\,\ref{fig:ESR4}(a) together with the ESR gap $\Delta$. It must be noted that both $\textit{H}_{\rm C}$ and $\Delta$ are determined by the same residual anisotropies, which pin a certain spatial direction of the transversal net moments in the crystal \cite{42}. The gap $\Delta$ is still visible in ESR far above $T_{2}$ up to $T > T_{1}$ owing to the dynamic nature of this spin probe.

The temperature dependence of the $\textit{g}$-factor, though exhibiting a substantial scatter, reveals a tendency to decrease with decreasing temperature and crosses the value $\textit{g}$ = 2 which corresponds to the spin-only $\textit{g}$-factor [Fig.\,\ref{fig:ESR4}(b)]. Deviations of the $\textit{g}$-factor of Cu$^{2+}$ and Ir$^{4+}$ from $\textit{g}$ = 2 are expected due to a combined action of the spin-orbit coupling and the lower symmetry components of the electrical ligand field \cite{41}. The latter arises due to a distortion of the oxygen ligand octahedron. As discussed above, La$_{2}$CuIrO$_{6}$ possesses a low-symmetry crystal structure with triclinic $\textit{P}\overline{1}$ space group. The structural refinement reveals two non-equivalent octahedra of both CuO$_{6}$ and IrO$_{6}$. All bonds between the central metal ions and oxygen ligands in the respective octahedra are of different length. However, it is possible to identify the direction of the strongest distortion, which can be defined for all octahedra as an elongation axis. In the two CuO$_{6}$ octahedra this axis is parallel to the Cu1 -- O1 bond of 2.159 \r{A}, and Cu2 -- O2 bond of 2.185 \r{A}, respectively [Fig.\,\ref{fig:XRD}(c)]. In the case of elongation, the values of the g-factor tensor for Cu$^{2+}$ lie typically in the range $g_{\Vert}$ = 2.15 -- 2.30 and $g_{\perp}$ = 2.01 -- 2.1 \cite{41}. Note that both $g_{\Vert}$ and $g_{\perp}$, as well as the powder average $g_{\rm av}$ = (1/3)$g_{\Vert}$ + (2/3)$g_{\perp}$, are larger than 2. In the case of IrO$_{6}$ octahedra, one of them is elongated along the bond Ir1 -- O1 of 2.075 \r{A}, whereas the elongation axis of the other octahedron is parallel to the bond Ir2 -- O2 of 2.088 \r{A} [Fig.\,\ref{fig:XRD}(c)]. 

The $\textit{g}$-factor tensor components for Ir$^{4+}$ fall in the range $g_{\Vert} < 2 < g_{\perp}$ in the case of elongation, with values of $g_{\Vert}$ ranging from 2 to 0, and of $g_{\perp}$ from 2 to 2.73 in extreme cases \cite{41}. The powder averaged $\textit{g}$-factor $\textit{g}_{\rm av}$ is expected to be slightly smaller than 2. Apparently, the ESR mode emerging below $\sim$ 113 K is dominated by the resonance of short-range ordered Cu spins with a precession frequency corresponding to $\textit{g}_{\rm av}$ slightly larger than 2. Note that the emerging signal is first detected at a resonance field corresponding to an even larger value of the effective $\textit{g}$-factor $\textit{g}_{\rm eff}$ = 2.08 [Fig.\,\ref{fig:ESR2}(c)]. Upon lowering the temperature across $T_{1}$, at which the static order is detected by $\mu$SR [Section H], the Ir spins get progressively more involved in the resonance due to the exchange coupling between the two spin subsystems. Consequently, the common precession frequency slides to the values corresponding to $\textit{g}_{\rm av}$ slightly smaller than 2. The observed temperature dependent shift of the resonance field $\textit{H}_{0}$ measured at a fixed excitation frequency [Fig.\,\ref{fig:ESR2}(c)] is determined by both the $\textit{T}$-dependence of the gap $\Delta$ and of the $\textit{g}$-factor (Fig.\,\ref{fig:ESR4}). 

\vspace{-5pt}\subsection{Neutron Depolarization Measurement}\vspace{-5pt}

\begin{figure}[t]
\includegraphics[width=1\columnwidth]{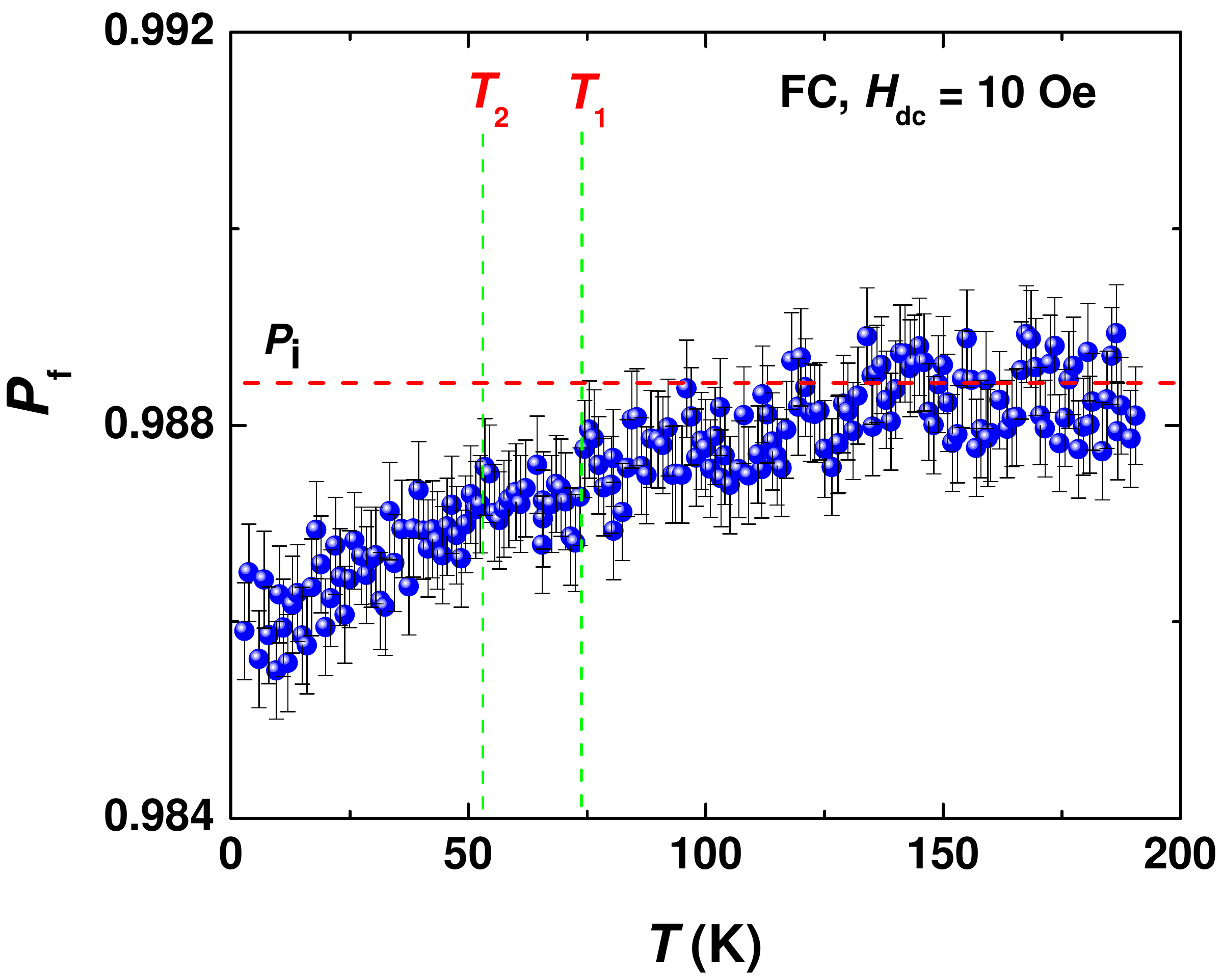}
\caption{Temperature dependence of the transmitted neutron polarization intensity for La$_{2}$CuIrO$_{6}$.}
\label{fig:NDP}\vspace{-5pt}
\end{figure}

The development of the uncompensated moments in the AFM short-range ordered regime in La$_{2}$CuIrO$_{6}$ is further corroborated by the neutron depolarization measurements. Here we pass a polarized neutron beam through the sample and record the change in polarization. First, the flipping ratio ($\textit{R}$) of the two types of incident neutron spins are measured from the respective transmitted intensities. Then the depolarization coefficient ($\textit{D}$) is calculated using \cite{50,51}:
\begin{equation}
R=\dfrac{1-P_{\rm i}DP_{\rm A}}{1+(2f-1)P_{\rm i}DP_{\rm A}},
\label{Eq:NDP}
\end{equation}
where $\textit{P}_{\rm i}$ is the incident neutron beam polarization, $\textit{P}_{\rm A}$ is the efficiency of analyzer crystal and $\textit{f}$ is the efficiency of \textit{dc} spin filter. Subsequently, the transmitted neutron beam polarization ($\textit{P}_{\rm f}$) is measured using, $P_{\rm f} = DP_{\rm i}$. Here the measurement is performed in FC configuration with $\textit{H}_{\rm dc}$ = 10 Oe and the results are displayed in Fig.\,\ref{fig:NDP}. When a polarized neutron beam passes through a ferromagnetic domain, the dipolar field induces Larmor precession of the neutron spins and depolarizes the neutron beam. So, for obvious reason no depolarization occurs in paramagnet and AFM compounds.

From the temperature dependent depolarization study of La$_{2}$CuIrO$_{6}$ (Fig.\,\ref{fig:NDP}), it is apparent that a weak but finite depolarization happens below $\sim$ 115 K. A slight enhancement of the depolarization below $\sim T_2$ is also evident. Here the data are normalized to the efficiency of the polariser (0.9883). The finite depolarization indicates that the presence of a weak intrinsic ferromagnetism in La$_{2}$CuIrO$_{6}$ that starts to grow up at a temperature below $\sim$ 115 K. With decreasing temperature, the FM-like contribution keeps on increasing and gets enhanced below $\sim T_2$. This is in agreement with the ESR and magnetometry measurements. Compared to the conventional FM materials \cite{52,53,54}, the depolarization effect is weak in the present case as it originates from the uncompensated AFM spin configuration of La$_{2}$CuIrO$_{6}$.

\vspace{-5pt}\subsection{NMR Measurements}\vspace{-5pt}

\begin{figure}[t]
\includegraphics[width=1\columnwidth]{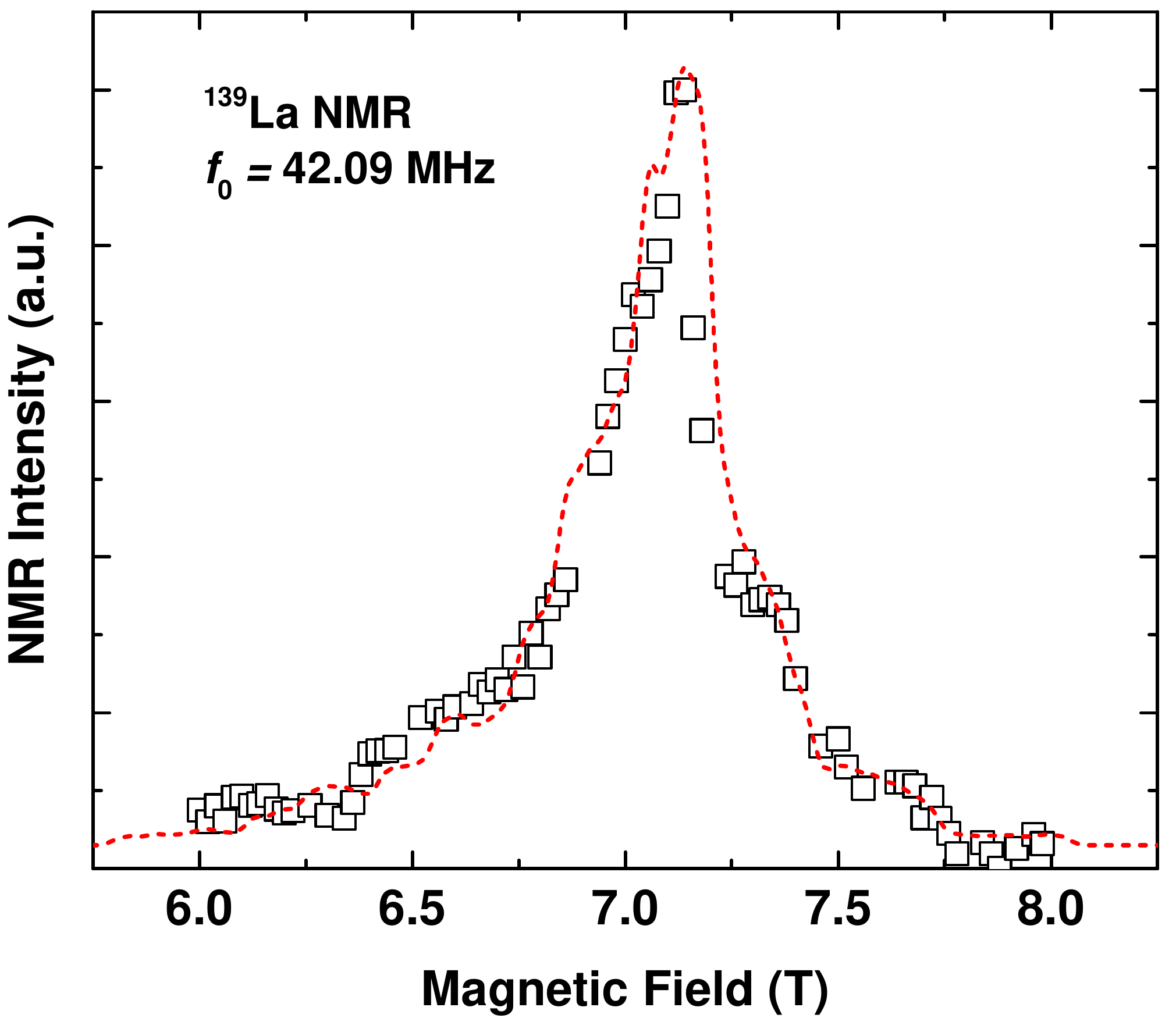}
\caption{Experimental $^{139}$La NMR field sweep spectrum at $\textit{T} = 150$\,K. The red dashed line shows the result of modeling as detailed in the text.}
\label{fig:NMR1}\vspace{-5pt}
\end{figure}

The examinations of the complex magnetic behavior in La$_{2}$CuIrO$_{6}$ by means of ESR and neutron scattering have been complemented by $^{139}$La--NMR studies to further elucidate the static and low frequency dynamic magnetic properties of the system. 
$^{139}$La is an NMR-active nucleus that has a natural abundance $\sim$ 100\% with the gyromagnetic ratio $\gamma = 6.014$ MHz/T, a nuclear spin $\textit{I}$ = 7/2 and an electric quadrupole moment ($\textit{Q}$ = 20 fm$^{2}$). The Hamiltonian determining the NMR spectra is represented as \cite{55}: 
\begin{equation}
H=H_{Z}+H_{\rm n-e}+H_{\rm Q},
\label{Eq:NMR1}
\end{equation}
where the $\textit{H}_{Z}$ term represents the Zeeman interaction with the external magnetic field, $\textit{H}_{\rm Q}$ is the quadrupolar coupling and $\textit{H}_{\rm n-e}$ is an anisotropic transferred hyperfine coupling of the nuclear spin with the electron spin. The whole NMR spectrum of La$_{2}$CuIrO$_{6}$ can be reasonably well modeled according to equation\,\ref{Eq:NMR1} with powder averaging. A comparison of the calculated and experimental spectra at $\textit{T}$ = 150 K is shown in Fig.\,\ref{fig:NMR1}. Here the quadrupolar coupling constant $\nu_{\rm Q}$= 1.9 MHz and the asymmetry parameter $\eta_{\rm Q}$ = 0.89 were estimated within the point charge model, and hyperfine coupling tensor's main components and linewidth are the fitting parameters.

\begin{figure}[b]
\includegraphics[width=1\columnwidth]{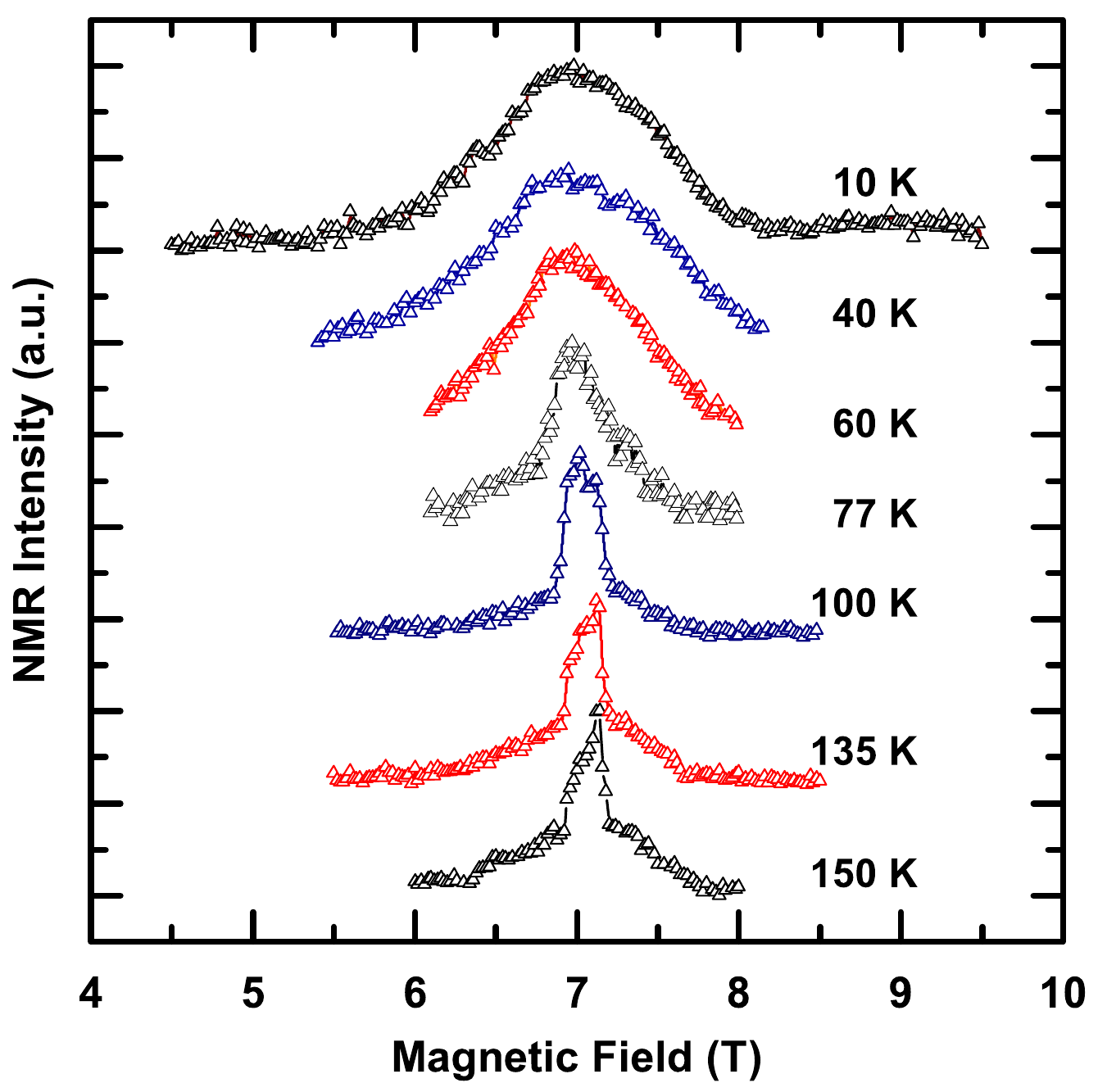}
\caption{Temperature evolution of the $^{139}$La field-swept NMR spectra.}
\label{fig:NMR2}\vspace{-5pt}
\end{figure}

Fig.\,\ref{fig:NMR2} presents the field-swept $^{139}$La NMR spectra obtained in the temperature range 10 K $\leq \textit{T} \leq$ 150 K. With decreasing temperature, the spectrum changes significantly. It starts to broaden above the magnetic phase transition, which could be a signature of the development of quasi-static electron spin correlations on the time-scale of NMR, consistent with the collective resonance mode seen by the ESR. The line continues to broaden below 75 K at least down to 10 K.  

\begin{figure}[b]
\includegraphics[width=1\columnwidth]{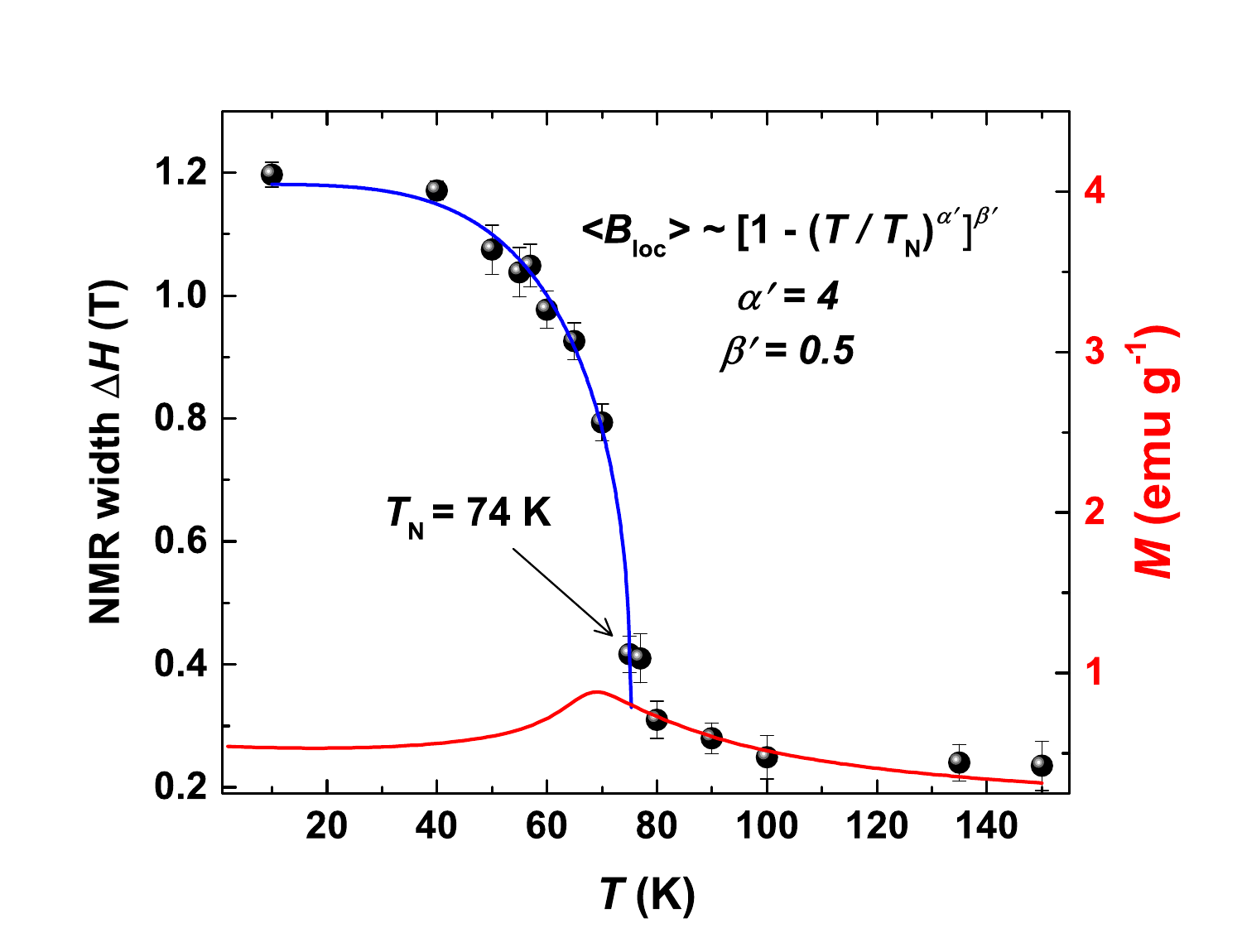}
\caption{Temperature dependence of the linewidth (symbols). The red solid line is the bulk static magnetization measured at the field 5 T; the blue line is a fit to the phenomenological function $\bigl[1-(T/T_{\rm N})^{\alpha'}\bigr]^{\beta'}$, as described in the text. }
\label{fig:NMR3}\vspace{-5pt}
\end{figure}

Usually in the magnetically ordered state the positions of the nuclei become magnetically non-equivalent and sense different static local fields. This results in the shift and splitting of the NMR spectrum. In the particular case of the present NMR experiment the inhomogeneous broadening is very large even in the magnetically ordered state, and the spectral structure is completely unresolved. Therefore, as a characteristic of the magnetically ordered state we have chosen the line width at the half-height of the NMR spectrum bearing in mind that it reflects both the magnitude of the local internal field on the $^{139}$La positions and their spatial distribution. The temperature dependence of the linewidth is presented in Fig.\,\ref{fig:NMR3}.

In the paramagnetic regime at $74\,\rm {K} \leqslant \textit{T} \leqslant 150$\,K, the linewidth follows the bulk static magnetization, as expected. However, at $T_{\rm N}$ the width increases step-like and deviates from the bulk magnetization. Such a dramatic broadening can be ascribed to the development of a specific distribution of internal fields in the magnetically ordered phase. Interestingly, the $\textit{T}$-dependence of the linewidth (Fig.\,\ref{fig:NMR3}) can be well fitted by the same phenomenological function $\bigl[1-(T/T_{\rm N})^{\alpha'}\bigr]^{\beta'}$ with $\alpha'$ = 4, $\beta'$ = 0.5, and $T_{\rm N} = T_{1} = 74$\,K as in the $\mu$SR-experiment (see below) [Fig.\,\ref{fig:uSR2}(a)] \cite{56}. Moreover, below $T_{\rm N}$ the NMR line broadening coincides with the temperature dependence of the intensity of the collective resonance mode observed in the ESR experiment (Fig.\,\ref{fig:ESR2}). The latter is determined by the magnitude of the precessing magnetization that can be considered as the magnetic order parameter. Therefore the local static magnetic fields probed by the NMR linewidth below 74 K are determined by the ordered part of the magnetic system. 

$\textit{NMR relaxation}$. Dynamic properties of the electron spin system were studied by measurements of the longitudinal $T_{\rm l}^{-1}$ and transversal relaxation $T_{\rm t}^{-1}$ rates at the maximum of the spectra in a temperature range from 150 K down to 40 K. Note that in magnetic substances both $T_{\rm l}^{-1}$ and $T_{\rm t}^{-1}$ are dominated by interactions with magnetic ions and probe fluctuating local magnetic fields perpendicular and parallel to the external magnetic field direction, respectively. The spectra rapidly broaden with lowering the temperature, yielding an overlap of the central line and the satellites of the spectrum. To determine the relaxation rates we have used a simple phenomenological description of the nuclear spin-echo intensity decay in the form \cite{57}:
\begin{equation}
M(\tau)=A_{1}e^{-(\tau/T_{\rm l})^{p}}+n_{0}
\label{Eq:NMR2}
\end{equation}
and
\begin{equation}
M(\tau)=A_{2}e^{-(\tau/T_{\rm t})}+n_{0}.
\label{Eq:NMR3}
\end{equation}
Here A$_{i}$ is the amplitude, $\textit{p}$ is a stretching parameter accounting for a distribution of spin lattice relaxation times and $\textit{n}_{0}$ accounts for the noise level. This reflects a distribution of fluctuation frequencies of the localized electron spins. Another origin of the stretched behavior of $\textit{M}(\tau)$ could be due to an excitation of only a part of the broad quadrupole split spectrum components in the NMR experiment. With increasing the local field magnitude and its distribution, the overlapping of the main component and of the quadrupole satellites increases resulting in the change of the $\textit{p}$ value.  

\begin{figure}[t]
\includegraphics[width=1\columnwidth]{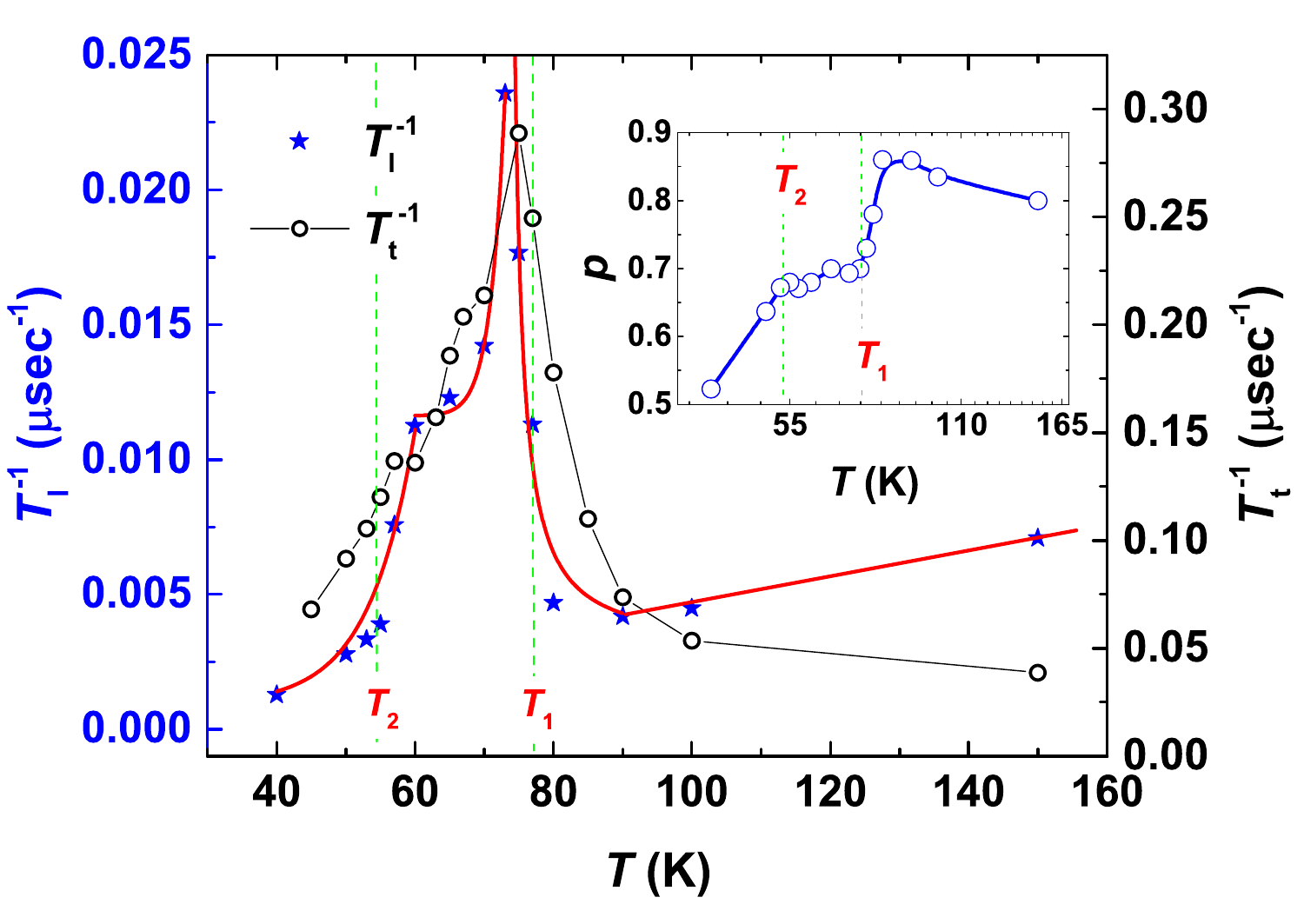}
\caption{Temperature dependence of the $^{139}$La longitudinal $T_{\rm l}^{-1}$ (filled stars) and transversal $T_{\rm t}^{-1}$ (circles) relaxation rates. Inset: stretching exponent $\textit{p}$ as a function of temperature. Solid lines are guides for the eyes.}
\label{fig:NMR4}\vspace{-5pt}
\end{figure}

A very steep rise of $T_{\rm l}^{-1}$ by approaching $T_{\rm N}$ from high temperature (shown in Fig.\,\ref{fig:NMR4}) proves that the compound undergoes a magnetic phase transition, and does not show a smooth crossover to a static, ordered state. The peak of $T_{\rm l}^{-1}$ occurs at $\sim T_{\rm N}$. However, below the peak the behavior of the relaxation rate is rather nontrivial. The $\textit{T}$-dependence of $T_{\rm l}^{-1}$ exhibits a shoulder in the temperature range between $T_{\rm N}$ and $T_2 = 54$\,K in the magnetization measurements. It is reasonable to associate this feature with the slowing down of the residual dynamics of small non-parallel components of copper and iridium sublattice magnetizations seen by ESR. The peculiarity at $T_2 = 54$\,K shows the freezing of such kind of fluctuations.

This scenario is further supported by the temperature dependence of the stretched exponent coefficient $\textit{p}$ (Fig.\,\ref{fig:NMR4}, inset). By approaching $T_{1}$ the coefficient $\textit{p}$ rapidly drops signifying the occurrence of magnetic order in the Cu spin system. By further decreasing the temperature, \textit{p} slightly decreases and drops again below $T_{2}$ reflecting a change of the local dynamic field distribution. The $\textit{T}$-dependence of $T_{\rm t}^{-1}$ is similar to that of $T_{\rm l}^{-1}$ (Fig.\,\ref{fig:NMR4}). In particular, it also exhibits a shoulder at $T_{2} < \textit{T} < T_{1}$ evidencing the occurrence of the fluctuating local fields parallel to the direction of the external magnetic field. They are not seen in the behavior of the linewidth and the lineshift due to a dominant contribution of the static internal fields from the ordered spins.

\begin{figure}[b]
\includegraphics[width=1\columnwidth]{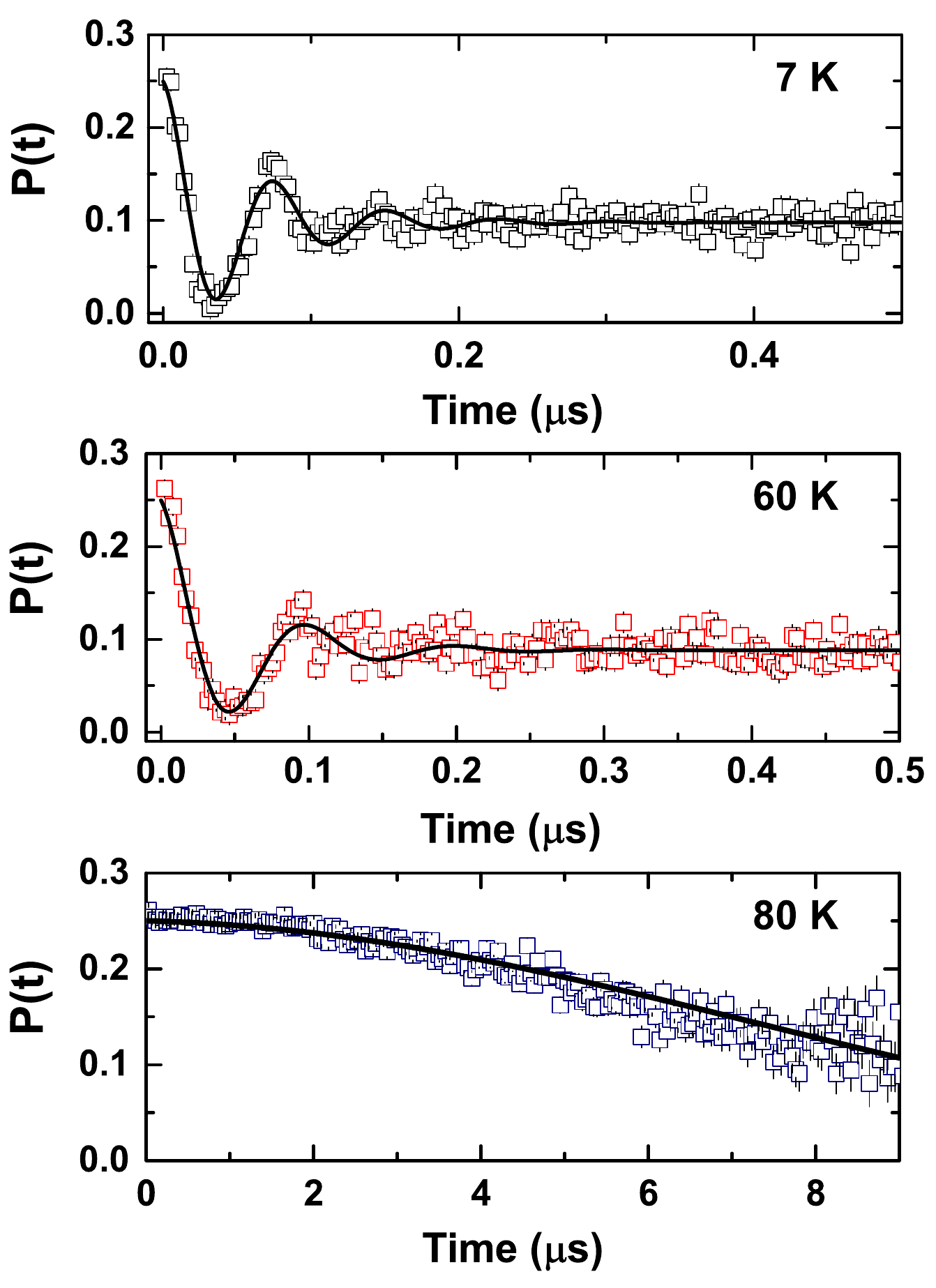}
\caption{Zero field $\mu$SR spectra (muon spin polarization versus time) at representative temperatures. The lines represent the theoretical description as detailed in the text.}
\label{fig:uSR1}\vspace{-5pt}
\end{figure}

\vspace{-5pt}\subsection{Zero field $\mu$SR Measurements}\vspace{-5pt}

We performed zero field $\mu$SR experiments to examine the magnetic order parameter and the spin dynamics in La$_{2}$CuIrO$_{6}$.  The time evolution of the muon spin polarization is shown in Fig.\,\ref{fig:uSR1}. In the high-temperature paramagnetic region, a weak Gaussian--Kubo--Toyabe relaxation of the signal is observed due to the dipole--dipole interaction of the muon magnetic moment with randomly oriented nuclear magnetic moments. With lowering the temperature below $T_{1}\sim 74$\,K, a spontaneous oscillation of the muon spin polarization is observed. This proves the development of the static internal field associated with long-range magnetic ordering. This is in agreement with the magnetic transition observed in the macroscopic measurements such as \textit{dc} as well as \textit{ac} magnetization and heat capacity. A well-defined $\mu$SR frequency below $\sim 74$\,K indicates the development of a distinct internal magnetic field at the muon site. We can, therefore, rule out the possibilities of a spin-glass state or an incommensurate state, confirming a commensurate magnetic order instead. The time dependence of the muon spin polarization is described by the following two-component functional form,
\begin{equation}
P(t)=\dfrac{2}{3}e^{-\lambda_{T}}\cos(2\pi f_{\mu}t+\theta)+\dfrac{1}{3}e^{-\lambda_{L}}
\label{Eq:uSR1}
\end{equation}
The 2/3 oscillating and 1/3 non-oscillating $\mu$SR signal fractions originate from the spatial averaging in powder samples, where 2/3 of the magnetic field components are perpendicular to the muon spin and cause a precession, while the 1/3 longitudinal field components do not. The relaxation of the oscillation, $\lambda_{\rm T}$, is a measure of the width of the static Gaussian field distribution $\Delta H = \lambda_{\rm T}/\gamma_{\mu}$. Dynamical effects are also present in $\lambda_{\rm T}$ while the relaxation of the second term, $\lambda_{\rm L}$, is due to dynamic magnetic fluctuations only. 

\begin{figure}[t]
\includegraphics[width=1\columnwidth]{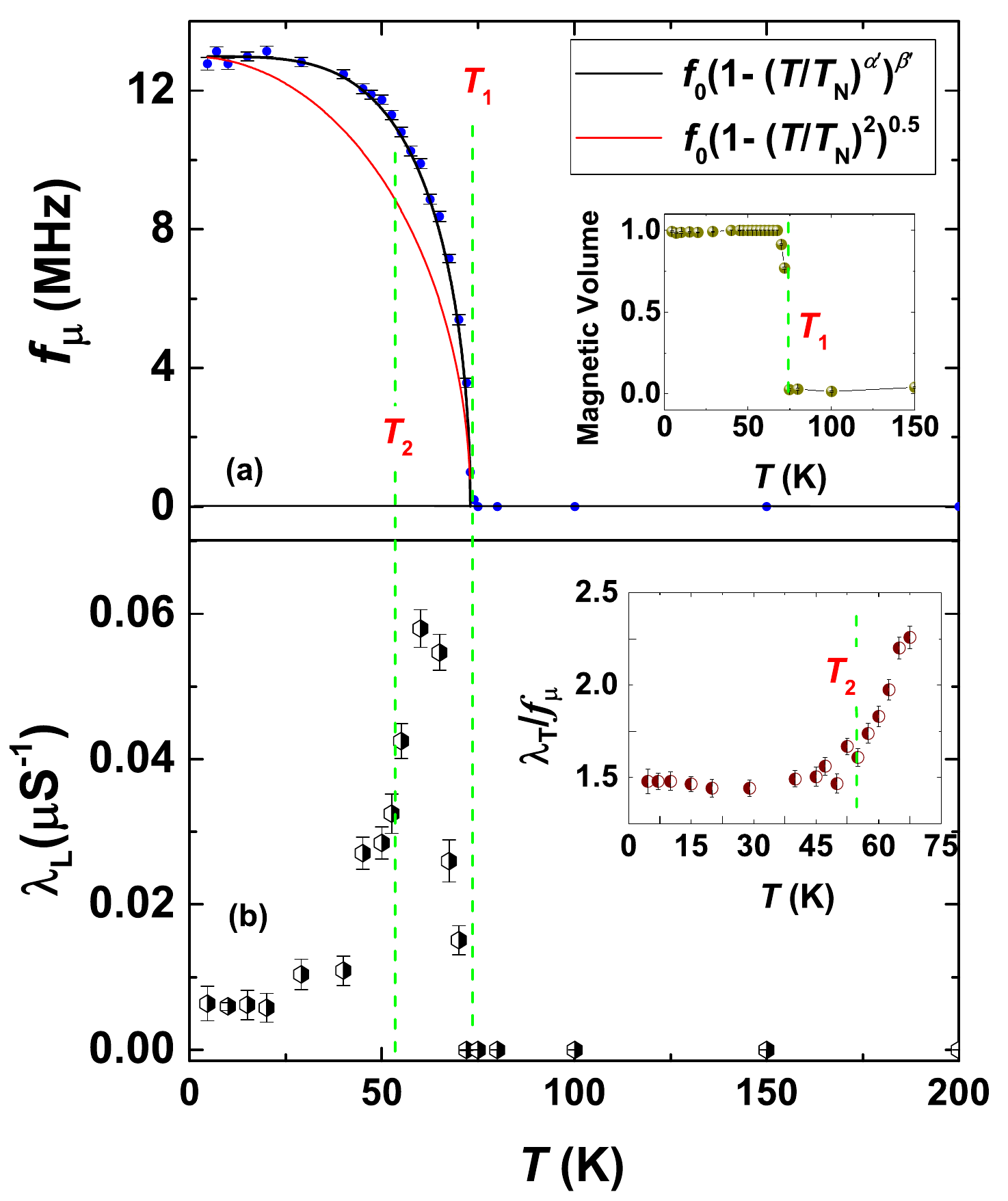}
\caption{(a) Temperature dependence of the muon spin precession frequency in La$_{2}$CuIrO$_{6}$ (main panel) and of the magnetic volume fraction (inset). Lines indicate the fit with different phenomenological models. (b) Temperature dependence of the muon spin lattice relaxation rate $\lambda_{\rm L}$(main panel) and of the normalized static line width $\lambda_{\rm T}$/$\textit{f}_{\mu}$ (inset).}
\label{fig:uSR2}\vspace{-5pt}
\end{figure}

The temperature dependence of the muon spin precession frequency ($\textit{f}_{\mu}$) and the magnetic volume fraction as obtained from the analysis are shown in Fig.\,\ref{fig:uSR2}(a) and its inset, respectively. The sudden increase of $\textit{f}_{\mu}$ below $\sim 74$ K represents a magnetic transition at this temperature. The magnetic volume fraction changes to $\sim 100$\% within a few degrees K indicating that the sample is homogeneous. The gradual increase of the $\mu$SR frequency, caused by the appearance of a static internal magnetic field at the muon site, proves a second order phase transition. From the measured $\mu$SR frequency value $\textit{f}_{\mu} = 12.98$ MHz at $T = 4.3$ K, a local internal field at the muon site of $\textit{H}_{\rm local} = 957.9$ Oe can be determined, using the relation $\textit{H}_{\rm local} = 2\pi f_\mu/\gamma _{\mu}$, where $\gamma_{\mu} = 2\pi \times 13.55$ kHz/Oe is the muon gyromagnetic ratio. Since $\textit{H}_{\rm local}$ is caused by superposition of contributions of dipole fields from nearby static Cu and Ir magnetic moments, $\textit{f}_{\mu}(\textit{T})$ is proportional to the average magnetic order parameter if no spin reorientations are present. The functional form of the temperature dependence of $\textit{f}_{\mu}$ cannot be described by a mean field behavior $\sim\sqrt{1-\bigl(T/T_{\rm N}}\bigr)^{2}$, where $T_{\rm N} = T_1$ is the Neel temperature of the AFM ordering. However, it can be described by the phenomenological function $\textit{f} = \textit{f}_{0}\bigl[1-(T/T_{\rm N})^{\alpha'}\bigr]^{\beta'}$, where $\alpha'$ and $\beta'$ are 4 and 0.5, respectively. Here, $\beta'$ is the critical exponent describing the behavior close to $T_{\rm N}$,  $\alpha'$ is used to adequately describe the data at low temperatures $T\rightarrow 0$. In mean field approximation $\alpha'$ is 2; simple spin wave theory for a ferromagnet gives $\alpha'$ = 3/2 (Bloch law). The larger value here reveals that low energy spin excitations are suppressed \cite{58,59}. 

The smooth increase of accurate data of $\textit{f}_{\mu}(\textit{T})$ below $T_{\rm N}$ without any kink or anomaly at 54 K proves that a spontaneous ordering of a magnetic sublattice, as indicated by the peak in the \textit{ac} susceptibility at this temperature, can only be associated with a very small moment size of less than 10$^{-2} \mu_{\rm B}$. Therefore, it is consistent with the freezing or long-range ordering of the weak spin canting induced by SOC in this compound.

In Fig.\,\ref{fig:uSR2}(b) we plot the $\mu$SR spin lattice relaxation rate $\lambda_{\rm L}$  as a function of temperature. A maximum of $\lambda_{\rm L}$ is found at $\sim 60$ K. Such a peak, in a magnetic system with a simple one-component order parameter associated with the magnetic fluctuation rate being equal to the muon spin precession frequency, is expected to be located very close to the magnetic ordering temperature $T_{\rm N} = 74$\,K, at which the spin correlation time diverges. However, in this system, we observe the onset of the strong increase of $\lambda_{\rm L}$ at $T_{\rm N}$. The reason for this observation is not clear yet. In La$_{2}$CuIrO$_{6}$, two different spin-spin correlations are found: strong longitudinal AFM correlations diverge at 75 K and weak transverse FM-like correlations diverge at 54 K. Both processes contribute to the spin lattice relaxation rate, however with different amplitudes $(\Delta B_{\mu})^{2}$ (form factors) at the site of the probe. The amplitude of the FM correlations, even though caused by small transverse magnetic moments of the order of 1/1000 $\mu_{\rm B}$ can have a similar absolute value as the amplitude of the strong AFM correlations. Therefore, in this system the superposition of two broad peaks may lead to the observed maximum in the muon spin lattice relaxation at 60 K. 

In the inset of Fig.\,\ref{fig:uSR2}(b) we plot the static line width $\lambda_{\rm T}$  normalized by the muon spin precession frequency $\textit{f}_{\mu}$. This is a measure of the homogeneity of the magnetic ordering. A constant value below 50 K proves a homogenous-system state without changes of the magnetic sublattice below this temperature. Above 50 K the ratio $\lambda_{\rm T}$/$\textit{f}_{\mu}$ is increasing towards $T_{\rm N}$. This is often found in magnetic systems close to the magnetic transition temperature since the magnetic coherence length is reduced close to $T_{\rm N}$ which gives rise to an increased static disorder. In La$_{2}$CuIrO$_{6}$ also a disordered state of transverse canted moments above 54 K may contribute to the increased static disorder.

\vspace{-5pt}\section{Discussion}\vspace{-5pt}

The occurrence of a small spontaneous ferromagnetic magnetization in antiferromagnets with the canted spin structure has been first understood in the pioneering works by Dzyaloshinsky and Moriya (DM) \cite{60,61}. Unlike in ferromagnets, this small magnetization is not related to the ferromagnetic exchange interaction between the spins. It is caused by the special antisymmetric anisotropic DM exchange interaction which arises due to the relativistic spin--orbit coupling in the absence of the inversion symmetry between the spins and adds to the usual antiferromagnetic Heisenberg exchange Hamiltonian. The DM interaction yields a small canting of the otherwise antiparallel spin sublattices. The resulting net ``FM--like'' transversal moment is usually just a fraction of the total spin moment. 

The experimental results obtained in the present work give strong arguments of a non-collinear antiferromagnetism of La$_{2}$CuIrO$_{6}$ [one of the possible spin canting scenarios is sketched in Fig.\,\ref{fig:XRD}(c)]. Though this canting could not be directly observed in the magnetic neutron diffraction, it is obviously due to the smallness of the associated net moment of the order of $\sim10^{-3} \mu_{\rm B}$. However, these small moments are responsible for rather peculiar magnetic properties of La$_{2}$CuIrO$_{6}$. The weak ferromagnetism becomes visible already in the dynamic regime above $T_1 = 74$\,K giving rise to the neutron spin depolarization and a collective ESR mode of the oscillating net moments. Magnetic neutron diffraction and $\mu$SR data evidence static AFM order of both Cu and Ir sublattices at $T_{1}$. However, it appears that the transverse net moments still exhibit a certain dynamical behavior below $T_{1}$ visible in the NMR and $\mu$SR relaxation rates. It can be understood as small incoherent fluctuations of the ordered sublattices around their equilibrium (antiferromagnetic) positions. Finally at $T_2 = 54$\,K, the net moments get statically locked giving rise to a weak anomaly in the magnetization at small fields. 

The multi-facet role of the spin--orbit coupling occurring at the different energy scales in La$_{2}$CuIrO$_{6}$ should be pointed out here: (i) the strong intra-ionic SOC of Ir$^{4+}$ of the order 0.5 eV ($\sim $6 000 K) couples the spin and orbital momentum giving rise to an effective $\textit{j}$ =1/2 ground state of Ir$^{4+}$ ions; (ii) the SOC of both Ir$^{4+}$ and Cu$^{2+}$ eventually selects specific spatial spin directions in the magnetically ordered state at $T_1 = 74$\,K via symmetric corrections to the Heisenberg superexchange (the pure Heisenberg-only isotropic spins would be insensitive to the lattice directions); (iii) due to the very low crystallographic symmetry of La$_{2}$CuIrO$_{6}$ that allows the antisymmetric DM interaction SOC finally yields weak transversal (weak FM) moments that get static at $T_2 = 54$\,K due to a residual magnetic anisotropy.

Currently, it is not possible to further detail the spin structure of La$_{2}$CuIrO$_{6}$. In particular, it is not clear if both Cu and Ir spin sublattices or only one of them are tilted and into which direction. The sketch in Fig.\,\ref{fig:XRD}(c) visualizes one of the possible spin patterns compatible with our data. All spins are lying in the $ac$-plane, within each layer the structure is almost collinear, and the spins in two layers are orthogonal to each other. The spin canting is strongly exaggerated in the sketch. It is assumed that both Ir and Cu sublattices are canted. This is likely to be the case since the DM interaction should act both within and between the different sublattices.

\vspace{-5pt}\section{Conclusions}\vspace{-5pt}

In this work, we presented the structural, magnetic and thermodynamic properties of the double perovskite La$_{2}$CuIrO$_{6}$ comprising interpenetrating and interacting sublattices of the Cu$^{2+}$ spins $s = 1/2$ and Ir$^{4+}$ pseudospins $\textit{j}$ = 1/2. The comprehensive analysis of XRD and neutron diffraction patterns reveals that the crystal structure of La$_{2}$CuIrO$_{6}$ is indeed triclinic $\textit{P}\overline{1}$ in contrast to the reported monoclinic $\textit{P}2_{1}/\textit{n}$. The \textit{dc} magnetization studies reveal an antiferromagnetic transition at $T_{1} = 74$\,K and the formation of a small uncompensated moment at even lower temperatures. The occurrence of the magnetic phase transition at $T_{1}$ is confirmed by the specific heat measurements. \textit{ac} susceptibility measurements prove a weak dynamic anomaly at $T_2 = 54$\,K which is suppressed in applied \textit{dc} fields. A Curie--Weiss fit of the inverse \textit{dc} susceptibility yields $\theta_{\rm CW} = + 33.7$\,K and the hysteresis loop opens up below $T_{2}$. The temperature dependent neutron diffraction and $\mu$SR data analysis assigns $T_{1}$ as the onset of AFM ordering with a commensurate magnetic structure. The spin configuration might be of a non-collinear type with an orthogonal orientation of magnetic moments in the neighboring AFM planes stacked along the crystallographic $\textit{b}$ axis. A smooth development of a static muon spin precession frequency proves a common magnetic ordering of strongly coupled Cu spins and Ir pseudospins.  The specific heat, ESR and neutron depolarization studies reveal that short-range magnetic correlations appear at much higher temperature than $T_{1}$. In particular, a crossover from a purely paramagnetic resonance regime to the development of a collective resonance mode in the short range ordered state of La$_{2}$CuIrO$_{6}$ is observed by ESR for $T_{1} \leqslant \textit{T} \leqslant 113$\,K. Very interestingly, $\mu$SR and NMR relaxation measurements reveal a residual spin dynamics at $T_{2} < T < T_{1}$.

It follows from the analysis of our experimental data that the peculiar magnetic properties of La$_{2}$CuIrO$_{6}$ are caused by an intimate interplay of the two coupled sublattices of Cu$^{2+}$ spins $s = 1/2$ and Ir$^{4+}$ pseudospins $j = 1/2$, respectively. Here, a combined effect of the spin-orbit coupling and of the low-symmetry crystallographic structure yields a small canting of the ordered spins. Interestingly, the resulting transversal moments exhibit low-frequency dynamics in a broad temperature range below $T_{1} = 74$ K suggesting that the antiferromagnetic sublattices still slightly fluctuate. Eventually the spin system gets fully statically ordered below $T_{2}$ due to the locking of the transverse moments. Obviously, our results call for further work on the present compound La$_{2}$CuIrO$_{6}$ and other $3d-5d$ mixed double perovskites to elucidate the impact of the interplay between the strong spin-orbit coupling and the crystal structure on the magnetism of this new class of complex transition oxides.

\vspace{-5pt}\section*{Acknowledgments}\vspace{-5pt}

We would like to thank S. M\"uller-Litvanyi, J. Werner and S. Ga\ss\, for technical support. D.\,S.\,I. acknowledges helpful discussions with J. Hunger and F. Damay. E.\,V. is grateful to G. Khaliullin for helpful discussions of spin dynamics in 5$\textit{d}$-perovskites. The PSI--$\mu$SR crew members are gratefully acknowledged for their support during the experiments. Funding support from the German Research Foundation (DFG) within projects KA 1694/8-1 (V.\,K.), WU595/3-3 (S.\,W.), WO1532/3-2 (A.\,U.\,B.\,W.) and within the collaborative research center SFB 1143, projects B01 (S.W. and B.B.), C02 (R.\,S. and H.\,H.\,K.), and C03 (Y.\,A.\,O. and D.\,S.\,I.) is gratefully acknowledged. The work (E.\,V. and M.\,I.) has been supported in part by the project RFBR 14-02-01194.


\begin{thebibliography}{}
\bibitem{Mott} D. Pesin and L. Balents, \href{http://www.nature.com/nphys/journal/v6/n5/abs/nphys1606.html} {\emph{Nat. Phys.} \textbf{6}, 376 (2010)}.
\bibitem{2} G. Jackeli and G. Khaliullin, \href{http://journals.aps.org/prl/abstract/10.1103/PhysRevLett.102.017205} {\emph{Phys. Rev. Lett. } \textbf{102}, 017205 (2009)}.
\bibitem{3} F. Wang and T. Senthil, \href{http://journals.aps.org/prl/abstract/10.1103/PhysRevLett.106.136402} {\emph{Phys. Rev. Lett. } \textbf{106}, 136402 (2011)}.
\bibitem{4} Xiangang Wan, A. M. Turner, A. Vishwanath, and S. Y. Savrasov, \href{http://journals.aps.org/prb/abstract/10.1103/PhysRevB.83.205101} {\emph{Phys. Rev. B} \textbf{83}, 205101 (2011)}.
\bibitem{5} L. Balents, \href{http://www.nature.com/nature/journal/v464/n7286/full/nature08917.html} {\emph{Nature (London)} \textbf{464}, 199 (2010)}.
\bibitem{6}J. Chaloupka, G. Jackeli, and G. Khaliullin, \href{http://journals.aps.org/prl/abstract/10.1103/PhysRevLett.105.027204} {\emph{Phys. Rev. Lett.} \textbf{105}, 027204 (2010)}.
\bibitem{7} B. J. Kim, H. Jin, S. J. Moon, J.-Y. Kim, B.-G. Park, C. S. Leem, J. Yu, T. W. Noh, C. Kim, S.-J. Oh, J. H. Park,V. Durairaj, G. Cao, and E. Rotenberg,  \href{http://journals.aps.org/prl/abstract/10.1103/PhysRevLett.101.076402} {\emph{Phys. Rev. Lett.} \textbf{101}, 076402 (2008)}.

\bibitem{8}	G. Demazeau, B. Siberchicot, S. Matar, C. Gayet, and A. Largeteau,  \href{http://scitation.aip.org/content/aip/journal/jap/75/9/10.1063/1.355909} {\emph{J. Appl. Phys.} \textbf{75}, 4617 (1994)}.
\bibitem{9}	R. C. Currie, J. F. Vente, E. Frikkee, and D. J. W. Ijdo, \href{http://www.sciencedirect.com/science/article/pii/S0022459685712029} {\emph{J. Solid State Chem.} \textbf{116}, 199 (1995)}.
\bibitem{10}	M. Uhl, S. Matar, and B. Siberchicot, \href{http://www.sciencedirect.com/science/article/pii/S030488539800122X} {\emph{J. Magn. Magn. Mater.} \textbf{187}, 201 (1998)}.
\bibitem{11}	G. Cao, A. Subedi, S. Calder, J.-Q. Yan, J. Yi, Z. Gai, L. Poudel, D. J. Singh, M. D. Lumsden, A. D. Christianson, B. C. Sales, and D. Mandrus, \href{http://journals.aps.org/prb/abstract/10.1103/PhysRevB.87.155136} {\emph{Phys. Rev. B } \textbf{87}, 155136 (2013)}.
\bibitem{12}	G. Blasse, \href{http://www.sciencedirect.com/science/article/pii/0022190265804092} {\emph{J. Inorg. Nucl. Chem. } \textbf{27}, 993 (1965)}.
\bibitem{13}	A.V. Powell, J. G. Gore, and P. D. Battle, \href{http://www.sciencedirect.com/science/article/pii/092583889390864J} {\emph{J. of Alloys and Comp. } \textbf{201}, 73 (1993)}.
\bibitem{14}	E. M. Ramos, I. Alvarez, M. L. Veiga, and C. Pico, \href{http://www.sciencedirect.com/science/article/pii/0025540894900086} {\emph{Mat. Res. Bul.} \textbf{29}, 881 (1994)}.
\bibitem{15}	J. Rodriguez-Carvajal, \href{http://www.sciencedirect.com/science/article/pii/092145269390108I} {\emph{Physica B } \textbf{192}, 55 (1993)}.
\bibitem{16}	T. Roisnel, and J. Rodriguez-Carvajal, \href{http://www.scientific.net/MSF.378-381.118} {\emph{Mater. Sci. Forum } \textbf{118}, 378 (2001)}.
\bibitem{17}	C. Golze, A. Alfonsov, R. Klingeler, B. Buchner, V. Kataev, C. Mennerich, H.-H. Klauss, M. Goiran, J.-M. Broto, H. Rakoto, S. Demeshko, G. Leibeling, and F. Meyer \href{http://journals.aps.org/prb/abstract/10.1103/PhysRevB.73.224403} {\emph{Phys. Rev. B } \textbf{73}, 224403 (2006)}.
\bibitem{18}	Pierre Villars, Material Phases Data System (MPDS), CH-6354 Vitznau, Switzerland (ed.), La$_{2}$CuIrO$_{6}$ (CuLa$_{2}$IrO$_{6}$) Crystal Structure, $sd-1211389$ \href{http://materials.springer.com/isp/crystallographic/docs/sd_1211389} {\emph{(Springer-Verlag GmbH, Heidelberg, 2014)}}.
\bibitem{19}	M. T. Anderson, K. B. Greenwood, G. A. Taylor, and K. R. Poeppelmeier,  \href{http://www.sciencedirect.com/science/article/pii/007967869390004B} {\emph{Prog. Sol. Stat. Chem.} \textbf{22}, 197 (1993)}.
\bibitem{20}	P. M. Woodward, \href{http://journals.iucr.org/b/issues/1997/01/00/br0052/br0052.pdf} {\emph{Acta Cryst. B} \textbf{53}, 32 (1997)}.
\bibitem{21}	C. J. Howard, B. J. Kennedy, and P. M. Woodward, \href{http://journals.iucr.org/b/issues/2003/04/00/ta5000/index.html} {\emph{Acta Cryst. B} \textbf{59}, 463 (2003)}.
\bibitem{22}	P. D. Battle, J. B. Goodenough, and R. Price, \href{http://www.sciencedirect.com/science/article/pii/0022459683901470} {\emph{J. Sol. State Chem.} \textbf{46}, 234 (1983)}.
\bibitem{23}	E. J. Cussen, D. R. Lynham, and J. Rogers, \href{http://pubs.acs.org/doi/abs/10.1021/cm0602388?journalCode=cmatex&quickLinkVolume=18&quickLinkPage=2855&selectedTab=citation&volume=18} {\emph{Chem. Mater.} \textbf{18}, 2855 (2006)}.
\bibitem{24}	C. D. Ling, B. J. Kennedy, Q. Zhou, J. R. Spencer, and M. Avdeev,  \href{http://www.sciencedirect.com/science/article/pii/S0022459610000319} {\emph{J. Solid State Chem.} \textbf{183}, 727 (2010)}.
\bibitem{25}	K. Manna, V. S. Bhadram, S. Elizabeth, C. Narayana, and P. S. Anil Kumar, \href{http://scitation.aip.org/content/aip/journal/jap/116/4/10.1063/1.4891015} {\emph{J. Appl. Phys.} \textbf{116}, 043903 (2014)}.
\bibitem{26}	D. Serrate, J. M. De Teresa, and M. R. Ibarra, \href{http://iopscience.iop.org/article/10.1088/0953-8984/19/2/023201} {\emph{J. Phys.: Condens. Matter} \textbf{19}, 023201 (2007)}.
\bibitem{27}	N. Narayanan, D. Mikhailova, A. Senyshyn, D. M. Trots, R. Laskowski, P. Blaha, K. Schwarz, H. Fuess, and H. Ehrenberg, \href{http://journals.aps.org/prb/abstract/10.1103/PhysRevB.82.024403} {\emph{Phys. Rev. B} \textbf{82}, 024403 (2010)}.
\bibitem{28}	K. Manna, D. Samal, S. Elizabeth, H. L. Bhat, and P. S. A. Kumar,\href{http://pubs.acs.org/doi/abs/10.1021/jp201206a?journalCode=jpccck&quickLinkVolume=115&quickLinkPage=13985&selectedTab=citation&volume=115}  {\emph{J. Phys. Chem. C } \textbf{115}, 13985 (2011)}.
\bibitem{29}	K. Manna, A. K. Bera, M. Jain, S. Elizabeth, S. M. Yusuf, and P. S. Anil Kumar,  \href{http://journals.aps.org/prb/abstract/10.1103/PhysRevB.91.224420} {\emph{Phys. Rev. B} \textbf{91}, 224420 (2015)}.
\bibitem{EndNote} Since the FM component is not completely suppressed for 10 kOe, this approximation will result in slightly reduced values for $M_{\rm FM}$. However, this is still a good approximation for the temperature dependence of $M_{\rm FM}$, since the FM component is nearly saturated for an external field of 10 kOe.
\bibitem{31}	A. U. B. Wolter, P. Wzietek, S. Sullow, F. J. Litterst, A. Honecker, W. Brenig, R. Feyerherm, and H.-H. Klauss \href{http://journals.aps.org/prl/abstract/10.1103/PhysRevLett.94.057204} {\emph{Phys. Rev. Lett.} \textbf{94}, 057204 (2005)}.
\bibitem{32}	M. R. Li, M. Retuerto, Z. Deng, T. Sarkar, J. Sanchez-Beniitez, M. C. Croft, T. S. Dasgupta, T. Das, T. A. Tyson, D. Walker, and M. Greenblatt, \href{http://pubs.acs.org/doi/abs/10.1021/cm503781s?journalCode=cmatex&quickLinkVolume=27&quickLinkPage=211&selectedTab=citation&volume=27} {\emph{Chem. Matter.} \textbf{27}, 211 (2015)}.
\bibitem{33}	W. K. Zhu, C.-K. Lu, W. Tong, J.M. Wang, H.D. Zhou, and S.X. Zhang, \href{http://journals.aps.org/prb/abstract/10.1103/PhysRevB.91.144408} {\emph{Phys. Rev. B} \textbf{91}, 144408 (2015)}.
\bibitem{34}	R. Morrow, R. Mishra, O. D. Restrepo, M. R. Ball, W. Windl, S. Wurmehl, U. Stockert, B. Buchner, and P. M. Woodward, \href{http://pubs.acs.org/doi/abs/10.1021/ja407342w} {\emph{J. Am. Chem. Soc.} \textbf{135}, 18824 (2013)}.
\bibitem{35}	K. Manna, S. Elizabeth, and P. S. A. Kumar, \href{http://scitation.aip.org/content/aip/journal/jap/119/4/10.1063/1.4940752} {\emph{J. Appl. Phys.} \textbf{119}, 043906 (2016)}.
\bibitem{36}	S. Yanez-Vilar, E. D. Mun, V. S. Zapf, B. G. Ueland, J. S. Gardner, J. D. Thompson, J. Singleton, M. Sanchez-Andujar, J. Mira, N. Biskup, M. A. Senaris-Rodriguez, and C. D. Batista, \href{http://journals.aps.org/prb/abstract/10.1103/PhysRevB.84.134427} {\emph{Phys. Rev. B} \textbf{84}, 134427 (2011)}.
\bibitem{37}	F. A. Lindemann, $\href{http://chemport.cas.org/cgi-bin/sdcgi?APP=ftslink&action=reflink&origin=npg&version=1.0&coi=1:CAS:528:DyaC3cXhsVCksQ%3D%3D&pissn=0028-0836&pyear=1990&md5=3feb5ed3a63dc71a1d39feee586a8d2b} {\emph{ Z. Phys.} \textbf{11}, 609 (1910)}.$
\bibitem{38}	J. W. Kim, Y. S. Oh, K. S. Suh, Y. D. Park, and K. H. Kim, \href{http://www.sciencedirect.com/science/article/pii/S0040603106006125} {\emph{Thermochimica Acta} \textbf{455}, 2 (2007)}.
\bibitem{39}	M. R. Lees, O. A. Petrenko, G. Balakrishnan, and D. McK. Paul, \href{http://journals.aps.org/prb/abstract/10.1103/PhysRevB.59.1298} {\emph{Phys. Rev. B} \textbf{59}, 1298 (1999)}.
\bibitem{40}	J. C. P. Poole, Electron spin resonance: A comprehensive treatise on experimental techniques (Dover Publications, Inc., Mineola, N.Y., 1996).
\bibitem{41}	A. Abragam and B. Bleaney, Electron paramagnetic resonance of transition ions (Clarendon Press, Oxford, 1970).
\bibitem{42}	E. A. Turov, Physical properties of magnetically ordered crystals, edited by A. Tybulewicz and S. Chomet (Academic press New York and London, 1965).
\bibitem{43}	B. Keimer, R. J. Birgeneau, A. Cassanho, Y. Endoh, M. Greven, M. A. Kastner, and G. Shirane, \href{http://link.springer.com/article/10.1007/BF01344067} {\emph{Z. Phys. B.} \textbf{91}, 373 (1993)}.
\bibitem{44}	S. Bahr, A. Alfonsov, G. Jackeli, G. Khaliullin, A. Matsumoto, T. Takayama, H. Takagi, B. Buchner, and V. Kataev, \href{http://journals.aps.org/prb/abstract/10.1103/PhysRevB.89.180401} {\emph{Phys. Rev. B} \textbf{89}, 180401 (2014)}.
\bibitem{45}	N. A. Bogdanov, V. M. Katukuri, J. Romhanyi, V. Yushankhai, V. Kataev, B. Buechner, J. van den Brink, and L. Hozoi, \href{http://www.nature.com/ncomms/2015/150624/ncomms8306/full/ncomms8306.html} {\emph{Nat. Commun.} \textbf{6}, 7306 (2015)}.
\bibitem{46}	J. Kim, A. H. Said, D. Casa, M. H. Upton, T. Gog, M. Daghofer, G. Jackeli, J. van den Brink, G. Khaliullin, and B. J. Kim, \href{http://journals.aps.org/prl/abstract/10.1103/PhysRevLett.109.157402} {\emph{Phys. Rev. Lett. } \textbf{109}, 157402 (2012)}.
\bibitem{47}	A. A. Aczel, A. M. Cook, T. J. Williams, S. Calder, A.D. Christianson, G.-X. Cao, D. Mandrus, Y. B. Kim, and A. Paramekanti, \href{http://journals.aps.org/prb/abstract/10.1103/PhysRevB.93.214426} {\emph{Phys. Rev. B} \textbf{93}, 214426 (2016)}.
\bibitem{48}	H. J. Fink, and D. Shaltiel, \href{http://journals.aps.org/pr/abstract/10.1103/PhysRev.130.627} {\emph{Phys. Rev.} \textbf{130}, 627 (1963)}.
\bibitem{49}	S. J. Williamson, and S. Foner, \href{http://journals.aps.org/pr/abstract/10.1103/PhysRev.136.A1102} {\emph{Phys. Rev.} \textbf{136}, A1102 (1964)}.
\bibitem{50}	S. M. Yusuf, and L. Madhav Rao, \href{http://link.springer.com/article/10.1007/BF02847303} {\emph{Pramana-J. Phys. } \textbf{47}, 171 (1996)}.
\bibitem{51}	K. Manna, D. Samal, A. K. Bera, S. Elizabeth, S. M. Yusuf, and P. S. A. Kumar, \href{http://iopscience.iop.org/article/10.1088/0953-8984/26/1/016002/meta} {\emph{J. Phys.: Condens. Matter } \textbf{26}, 016002 (2014)}.
\bibitem{52}	S. Mitsuda, H. Yoshizawa, and Y. Endoh, \href{http://journals.aps.org/prb/abstract/10.1103/PhysRevB.45.9788} {\emph{Phys. Rev. B} \textbf{45}, 9788 (1992)}.
\bibitem{53}	M. Halder, S. M. Yusuf, A. Kumar, A. K. Nigam, and L. Keller, \href{http://journals.aps.org/prb/abstract/10.1103/PhysRevB.84.094435} {\emph{Phys. Rev. B} \textbf{84}, 094435  (2011)}.
\bibitem{54}	S. M. Yusuf and L. Madhav Rao, \href{http://iopscience.iop.org/article/10.1088/0953-8984/7/29/016} {\emph{J. Phys.: Condens. Matter } \textbf{7}, 5891 (1995)}.
\bibitem{55}	A. Abragam, Principles of Nuclear Magnetism (Clarendon Press, Oxford, 1961).
\bibitem{56}	Malcolm F. Collins, Magnetic Critical Scattering (Oxford University Press, 1989).
\bibitem{57}	F. Bloch, \href{http://journals.aps.org/pr/abstract/10.1103/PhysRev.70.460} {\emph{Phys. Rev.} \textbf{70}, 460  (1946)}.
\bibitem{58}	R. Scheuermann, E. Roduner, G. Engelhardt, H.-H. Klauss, and D. Herlach,  \href{http://journals.aps.org/prb/abstract/10.1103/PhysRevB.66.144429} {\emph{Phys. Rev. B} \textbf{66}, 144429 (2002)}.
\bibitem{59}	R. Kubo, \href{http://journals.aps.org/pr/abstract/10.1103/PhysRev.87.568} {\emph{Phys. Rev.} \textbf{87}, 568 (1952)}.
\bibitem{60}I. Dzyaloshinsky, \href{http://www.sciencedirect.com/science/article/pii/0022369758900763} {\emph{J. Phys. Chem. Sol.} \textbf{4}, 241 (1958)}.
\bibitem{61} T. Moriya, \href{http://journals.aps.org/pr/abstract/10.1103/PhysRev.120.91} {\emph{Phys. Rev.} \textbf{120}, 91 (1960)}.


\end{thebibliography}
\end{document}